\documentclass[prb,twocolumn,groupedaddress,shownopacs,amssymb]{revtex4-2}
\usepackage{graphicx,psfrag,amsmath,amssymb,enumerate, overpic, subcaption, ragged2e, comment}
\usepackage{algpseudocode}

\usepackage[usenames, dvipsnames]{color}
\usepackage{bm}
\usepackage{hyperref}
\hypersetup{
    colorlinks=true,
    linkcolor=Blue,
    citecolor=Blue,
    filecolor=Blue,
    urlcolor=Blue,
    pdftitle={Vortex-core size and quantum geometry},
    }

\begin{document}

\title{Vortex-core size and quantum geometry in flat-band superconductors}
\author{M. \"Ozkurt and M. Iskin}
\affiliation{
Department of Physics, Ko\c{c} University, Rumelifeneri Yolu,
34450 Sar\i yer, Istanbul, T\"urkiye
}

\date{\today}

\begin{abstract}

We investigate the real-space vortex structure of a flat-band superconductor
described by the attractive Hubbard model on the Mielke checkerboard lattice.
Using a momentum-space mean-field analysis, we derive a closed-form expression
for the coherence length in terms of the quantum metric of the Bloch states and
a particle-hole-symmetric pair density. Self-consistent Bogoliubov-de Gennes
calculations on a finite disk confirm this prediction across a wide range of
fillings and interaction strengths. The coherence length is minimized at half
filling, diverges toward both band edges as the pair density vanishes, and
depends only logarithmically on the interaction strength. These features differ
qualitatively from the conventional BCS picture, which relies on a
well-defined Fermi surface and a kinetic mechanism for pair transport. Instead,
the vortex-core size is governed by a geometric pair mass arising from virtual
interband processes encoded in the quantum metric. Our results establish the
vortex core as a direct real-space manifestation of flat-band quantum geometry.

\end{abstract}

\maketitle

\section{Introduction}
\label{sec:intro}

Flat-band systems provide a unique setting for exploring
interaction-driven quantum phenomena. Because the kinetic
energy is strongly suppressed, interactions and the quantum
geometry of Bloch states can dominate the low-energy
physics, giving rise to unconventional forms of
superconductivity, superfluidity, and transport
properties~\cite{torma22, peotta23, yu25, liu25, gao25,
jiang25, verma26, kitamura26}. A central question in this
context is how superconducting correlations acquire a
characteristic spatial scale when the underlying band is
flat. In conventional Bardeen-Cooper-Schrieffer (BCS)
theory, the coherence length is controlled by the Fermi
velocity and therefore ultimately by band
dispersion~\cite{tinkham04, leggett}. In an exactly flat
band, however, neither a well-defined Fermi surface nor a
conventional kinetic-energy scale exists, leaving open the
question of what determines the size of spatial
inhomogeneities such as vortex cores.

Recent theoretical studies have identified quantum
geometry as a mechanism capable of replacing band
dispersion in flat-band superconductors
\cite{iskin23,chen23,thumin24,iskin24c,iskin25,li25,
virtanen25,lee25,xiao25,oh25,elden26,chen26}. Although
flat bands carry no conventional kinetic energy, virtual
interband processes encoded in the quantum metric generate
a finite effective mass for Cooper pairs. This geometric
pair mass produces a finite superfluid weight and a finite
healing length, establishing a direct connection between
Bloch-state geometry and macroscopic transport
properties. While this connection is now well established
for homogeneous systems~\cite{torma22, peotta23, yu25,
liu25, gao25, jiang25, verma26, kitamura26}, much less is
known about its manifestation in spatially inhomogeneous
situations. In particular, it remains unclear whether the
coherence length predicted from bulk quantum-geometric
theories survives in fully self-consistent vortex states,
where order-parameter suppression, density
redistribution, and finite-size effects become
important~\cite{li25}. Resolving this issue is essential
for connecting quantum geometry to experimentally
accessible real-space observables.

In this work, we address this problem by investigating a
singly quantized vortex in a flat-band superconductor
described by the attractive Hubbard model on the Mielke
checkerboard lattice~\cite{mielke91}. This lattice
contains an exactly flat band touching a dispersive band
through a quadratic band-touching point, providing a
minimal platform in which quantum-geometric effects are
both analytically transparent and numerically
accessible~\cite{iskin19}. We first develop a
momentum-space description and derive a closed-form
expression for the coherence length in terms of the
quantum metric and a particle-hole-symmetric pair
density. The resulting prediction contains no adjustable
parameters beyond the interaction strength, filling, and
an ultraviolet cutoff fixed independently through a
state-counting procedure. We then solve the
Bogoliubov-de Gennes (BdG) equations self-consistently on
a finite disk and extract the vortex-core size directly
from the spatial profile of the order parameter.

Our calculations reveal a remarkably close agreement
between the momentum-space prediction and the real-space
BdG results throughout the flat-band regime. The
coherence length is smallest near half filling and
diverges toward both band edges as the density of mobile
pairs vanishes. Moreover, its dependence on interaction
strength is only logarithmic, reflecting the logarithmic
enhancement of the geometric pair mass associated with
the quadratic band-touching point. The remaining
quantitative differences can be traced primarily to the
infrared cutoff imposed by the finite system size, which
limits the development of the asymptotic low-energy
behavior. These findings establish the vortex-core size
as a direct real-space manifestation of the
quantum-geometric pair mass and demonstrate that quantum
geometry can assume the role conventionally played by
band dispersion in determining the spatial structure of a
superconductor.

The remainder of this paper is organized as follows.
In Sec.~\ref{sec:kspace}, we introduce the momentum-space
formalism, including the lattice geometry, continuum model, 
mean-field theory, self-consistency equations, and
the resulting coherence-length expression. 
In Sec.~\ref{sec:bdg_form}, we develop the real-space BdG formalism 
and obtain self-consistent vortex solutions, from which we extract the 
vortex-core size and compare it with the momentum-space prediction. 
Finally, in Sec.~\ref{sec:conclusion}, we summarize our main findings. 
Technical details of the Bessel-function expansion and the BdG 
self-consistency relations are presented in 
Apps.~\ref{app:matrix}-\ref{app:observables}.

\section{Momentum-space formalism}
\label{sec:kspace}

Our analytical framework is based on the two-band Mielke checkerboard 
lattice~\cite{iskin19}. Its exactly flat lower band, together with the 
presence of both time-reversal and sublattice-exchange symmetries, makes 
it a particularly transparent platform for establishing the key concepts 
and deriving the closed-form results that form the foundation of the 
real-space BdG analysis.

\subsection{Mielke checkerboard lattice}
\label{sec:mielke_ham}

\begin{figure}[h!]
\centering
\begin{minipage}[b]{0.40\linewidth}
\centering
\includegraphics[width=\linewidth]{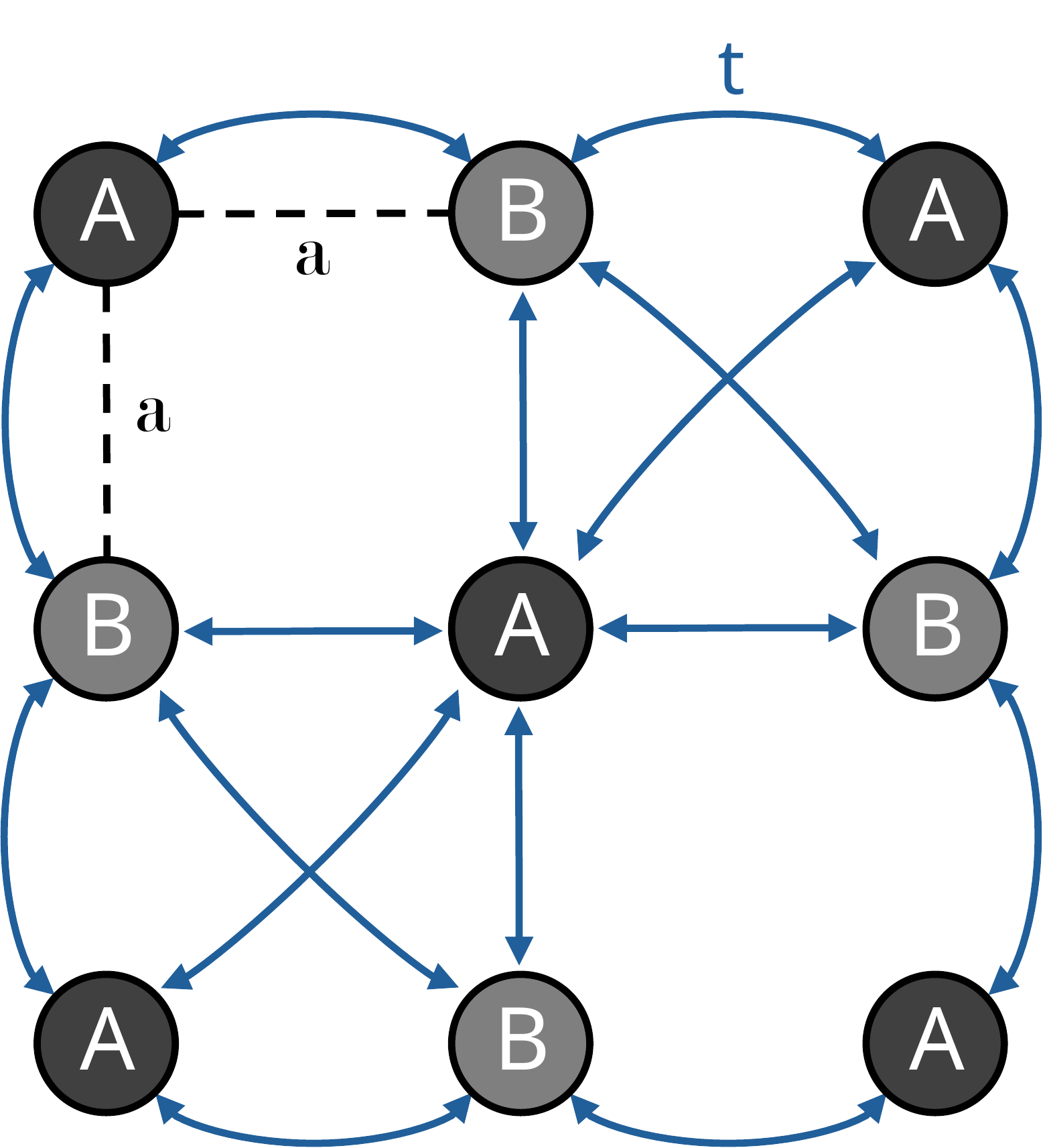}
\subcaption{}
\label{fig:lattice}
\end{minipage}
\hfill
\begin{minipage}[b]{0.58\linewidth}
\centering
\includegraphics[width=\linewidth]{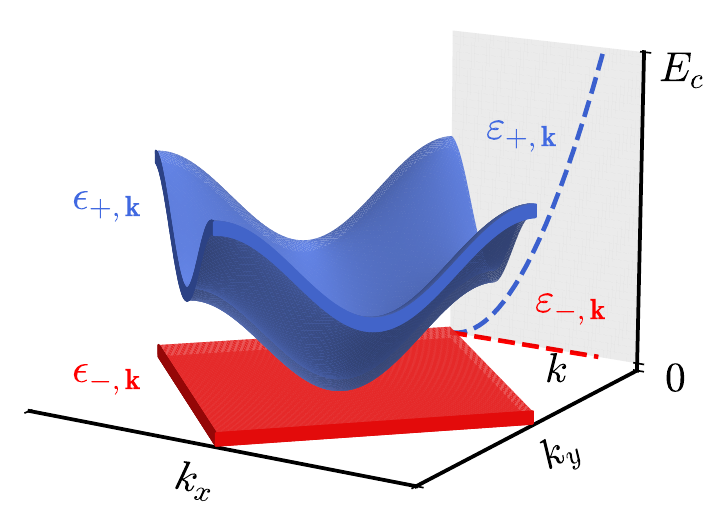}
\subcaption{}
\label{fig:band}
\end{minipage}
\caption{
\justifying
(a) Schematic of the two-orbital Mielke checkerboard lattice, 
with sublattices labeled $A$ and $B$. The lattice spacing and 
hopping amplitude are denoted by $a$ and $t$, respectively.
(b) Band structure of the lattice, showing the flat band (red) 
and the dispersive band (blue). The shaded region indicates the 
low-energy continuum sector retained in our analysis, bounded by 
the cutoff energy $E_c$ and projected onto the radial momentum
$k = |\bm{k}|$.
}
\label{fig:mielke_band_and_lattice}
\end{figure}

The Mielke checkerboard is a two-sublattice $(A,B)$ line-graph lattice 
with lattice spacing $a$ and hopping amplitude $t$, shown in
Fig.~\ref{fig:mielke_band_and_lattice}(a)~\cite{mielke91}. Its geometry 
enforces destructive interference between neighboring hopping paths, 
giving rise to an exactly flat Bloch band.
In the sublattice basis
$
\phi_{\sigma\bm{k}} = (c_{A\sigma\bm{k}}, c_{B\sigma\bm{k}})^\mathrm{T},
$
where $\sigma= \{\uparrow,\downarrow\}$ labels the spin and
$\bm{k}=(k_x,k_y)$ denotes the crystal momentum (with $\hbar=1$), the
noninteracting Hamiltonian is
$
\mathcal{H}_0 = \sum_{\sigma\bm{k}}\phi^{\dagger}_{\sigma\bm{k}}
H_{\bm{k}} \phi_{\sigma\bm{k}}.
$
The spin-independent Bloch Hamiltonian $H_{\bm{k}}$ can be written as
\begin{align}
H_{\bm{k}} = d_{\bm{k}}^{0} \tau_0 + d_{\bm{k}}^{x} \tau_x + d_{\bm{k}}^{z} \tau_z,
\end{align}
where $\tau_0$ is the identity matrix and $\tau_x$ and $\tau_z$ are
Pauli matrices acting on the sublattice degree of freedom. The
coefficients are
$
d_{\bm{k}}^{0} = 2t [1+\cos(k_x a)\cos(k_y a)],
$
$
d_{\bm{k}}^{x} = 2t [\cos(k_x a) + \cos(k_y a)],
$
and
$
d_{\bm{k}}^{z} = 2t\sin(k_x a)\sin(k_y a).
$
Since $\bm{d}_{\bm{k}}=\bm{d}_{-\bm{k}}$, the model possesses inversion
symmetry,
$
H_{\bm{k}}=H_{-\bm{k}}.
$
Furthermore, the absence of a $\tau_y$ component makes
$H_{\bm{k}}$ purely real, implying time-reversal symmetry,
$
H_{\bm{k}} = H^{*}_{-\bm{k}}.
$

The Bloch-band energies are
$
\epsilon_{s\bm{k}} = d_{\bm{k}}^{0} + s d_{\bm{k}},
$
where
$
d_{\bm{k}}
=
|\bm{d}_{\bm{k}}|
=
2t[1+\cos(k_xa)\cos(k_ya)]
$
and $s=\pm$. Since $d_{\bm{k}}^{0}=d_{\bm{k}}$, the spectrum reduces to
$
\epsilon_{-,\bm{k}} = 0
$
and
$
\epsilon_{+,\bm{k}}
=
4t[1+\cos(k_xa)\cos(k_ya)].
$
Thus, the lower band is perfectly flat, whereas the upper band is
dispersive, as shown in Fig.~\ref{fig:mielke_band_and_lattice}(b). The
two bands touch quadratically at the four symmetry-related points
$\bm{k}=(\pm\pi/a,0)$ and $\bm{k}=(0,\pm\pi/a)$, where
$d_{\bm{k}}=0$ and therefore $\epsilon_{s\bm{k}}=0$. Consequently, the
flat band is not separated from the dispersive band by an energy gap,
but instead forms a quadratic band touching.
Despite its minimal two-band structure, the Mielke checkerboard lattice
captures the essential ingredients of flat-band quantum geometry. The
quadratic touching points give rise to a singular quantum metric in the
long-wavelength limit, while the flat lower band suppresses conventional
kinetic contributions. As a result, this model provides an ideal setting
for isolating and analyzing geometric effects in pairing, superfluid
transport, and collective properties of flat-band systems.

\subsection{Effective continuum model}
\label{sec:low_energy}

Since the flat-band physics is governed by the vicinity of the quadratic
band-touching points, we formulate the remainder of this work within a
low-energy continuum description obtained by expanding $H_{\bm{k}}$ to
second order around one such point. Measuring energies relative to the
touching energy, the Bloch coefficients become
\begin{align}
d_{\bm{k}}^{0} = t a^2 k^2, \quad d_{\bm{k}}^{x} = t a^2 (k_x^2 - k_y^2), \quad
d_{\bm{k}}^{z} = 2 t a^2 k_x k_y,
\label{eq:lowE_dvec}
\end{align}
with $k^2 = k_x^2 + k_y^2$. Diagonalizing the resulting Hamiltonian yields
\begin{equation}
\varepsilon_{-,\bm{k}} = 0,
\qquad
\varepsilon_{+,\bm{k}} = 2 t a^2 k^2,
\label{eq:lowE_bands}
\end{equation}
so that the lower band remains perfectly flat at zero energy, while the
upper band acquires an isotropic quadratic dispersion. This behavior is
illustrated in the low-energy projection of
Fig.~\ref{fig:mielke_band_and_lattice}(b).

The geometric content of the Bloch states governing interband response
is encoded in the quantum metric tensor. For a two-band model, it is
convenient to express it in terms of the normalized vector
$
\hat{\bm{d}}_{\bm{k}}
= \frac{\bm{d}_{\bm{k}}}{|\bm{d}_{\bm{k}}|}
= \frac{1}{k^2} (k_x^2 - k_y^2, 0, 2 k_x k_y),
$
from which the quantum metric follows as
$
g^{\bm{k}}_{ij}
=
\frac{1}{2}
\partial_i \hat{\bm{d}}_{\bm{k}}
\cdot
\partial_j \hat{\bm{d}}_{\bm{k}},
$
leading to
\begin{equation}
g^{\bm{k}}_{ij}
=
\frac{2\delta_{ij}}{k^2}
-
\frac{2k_i k_j}{k^4},
\label{eq:metric}
\end{equation}
where $\partial_i \equiv \frac{\partial}{\partial k_i}$ and
$\delta_{ij}$ is the Kronecker delta.
In the following sections, we consider quantities such as the 
coherence length and effective inverse-mass tensors, which are
expressed as momentum sums of the quantum metric weighted by 
an isotropic function $w(k)$,
$
\mathcal{O}_{ij} = \sum_{\bm{k}} g^{\bm{k}}_{ij} w(k).
$
The off-diagonal component contains the factor
$k_x k_y w(k)/k^4$, which is odd under $k_x \to -k_x$ (or
$k_y \to -k_y$) and therefore vanishes upon Brillouin-zone integration.
Moreover, rotational symmetry enforces equality of the diagonal
components under $k_x \leftrightarrow k_y$. Consequently, all such
observables are diagonal and isotropic,
$
\mathcal{O}_{ij} = \mathcal{O}\delta_{ij},
$
so it suffices to consider a single diagonal component (e.g., $xx$),
allowing these quantities to be treated effectively as scalars.

\subsection{Mean-field Hamiltonian}
\label{sec:mf_ham}

We now include an on-site attractive Hubbard interaction of strength 
$U \ge 0$ between $\uparrow$ and $\downarrow$ fermions and treat it 
within the BCS mean-field approximation. Owing to the sublattice-exchange 
symmetry of the lattice, the superconducting order parameter is uniform 
in space~\cite{iskin19}, i.e, 
$
\Delta_A = \Delta_B \equiv \Delta_0,
$
which we take to be real without loss of generality.
Introducing the four-component Nambu spinor
$
\Phi_{\bm{k}} =
(
\phi_{\uparrow\bm{k}},
\phi^{\dagger}_{\downarrow,-\bm{k}}
)^{\mathrm T},
$
the mean-field Hamiltonian becomes
\begin{equation}
\mathcal{H}_{\mathrm{mf}}
=
\mathcal{E}_0 + \sum_{\bm{k}}
\Phi_{\bm{k}}^{\dagger}
\begin{pmatrix}
H_{\bm{k}} - \mu\tau_0 & \Delta_0\tau_0 \\[2pt]
\Delta_0\tau_0 & -H^{*}_{-\bm{k}} + \mu\tau_0
\end{pmatrix}
\Phi_{\bm{k}},
\label{eq:Hmf_kspace}
\end{equation}
where 
$
\mathcal{E}_0 = N\Delta_0^2/U + 2\sum_{\bm{k}} (d_{\bm{k}}^{0}-\mu)
$
is a constant energy offset, with $N$ the number of lattice sites, and
$\mu$ is the chemical potential.
Diagonalization yields the quasiparticle spectrum
\begin{equation}
E_{s\bm{k}}
=
\sqrt{\xi_{s\bm{k}}^2 + \Delta_0^2},
\qquad
\xi_{s\bm{k}} = \varepsilon_{s\bm{k}} - \mu,
\label{eq:qp_spectrum}
\end{equation}
which reflects the fact that uniform pairing does not mix the band 
index beyond the particle-hole structure. Consequently, the 
quasiparticle Hamiltonian remains block diagonal in the band index, 
which underlies the self-consistency equations derived in the 
following section.

\subsection{Self-consistency equations}
\label{sec:self_cons}

The pairing field and chemical potential are determined self-consistently by
requiring that the order parameter and particle number obtained from the
mean-field spectrum in Eq.~\eqref{eq:qp_spectrum} reproduce their assumed
values. The resulting gap and number equations can be written compactly 
as~\cite{iskin19}
\begin{align}
1 &= \frac{U}{2N}\sum_{s\bm{k}} \frac{\mathcal{X}_{s\bm{k}}}{E_{s\bm{k}}},
\label{eq:gap_eqn}\\
F &= 1 - \frac{1}{N}\sum_{s\bm{k}} \frac{\xi_{s\bm{k}}}{E_{s\bm{k}}}\mathcal{X}_{s\bm{k}},
\label{eq:number_eqn}
\end{align}
where
$
\mathcal{X}_{s\bm{k}} = \tanh \big[E_{s\bm{k}}/(2T)\big]
$
is the thermal factor (with $k_\mathrm{B} = 1$), and
$F=\mathcal{N}/N \in [0,2]$ denotes the filling per site, with $\mathcal{N}$
the total number of particles in the system. The BCS mean-field solution 
is obtained by simultaneously solving 
Eqs.~\eqref{eq:gap_eqn} and~\eqref{eq:number_eqn}
for $\Delta_0$ and $\mu$ at fixed $U$, $F$, and $T$~\cite{nsr85}.
To evaluate the momentum sums analytically, we set
$
\sum_{\bm{k}} \rightarrow \frac{A}{(2\pi)^2}\int_{|\bm{k}|<k_c} d^2k,
$
with $A=Na^2$ the total system area, which introduces an ultraviolet 
radial-momentum cutoff $k_c$.

To establish a direct correspondence with the real-space BdG calculations
performed on a disk of radius $R$, we fix the ultraviolet cutoff by
requiring that the number of states retained in the momentum-space
theory equals the number of unit cells contained within the 
corresponding real-space disk. Specifically, the number of continuum 
$\bm{k}$ states contained within a disk of radius $k_c$ is
$
N_c = \sum_{\bm{k}} 1 = \frac{A k_c^2}{4\pi},
$
whereas the number of unit cells within the corresponding lattice disk is
$
N_c = \frac{A}{2a^2},
$
since each unit cell occupies an area $2a^2$, with $N = 2N_c$ 
and $A=\pi R^2$. Matching these two expressions yields
$
k_c=\frac{\sqrt{2\pi}}{a},
$
which in turn determines the cutoff energy as
\begin{equation}
E_c = 2ta^2k_c^2 = 4\pi t,
\end{equation}
exceeding the bandwidth $8t$ of the original lattice model.
This prescription ensures that the continuum theory and the real-space
BdG formulation contain exactly the same number of single-particle
degrees of freedom. It therefore provides a natural bridge between the
bulk momentum-space calculations and the finite-size real-space
simulations discussed in Sec.~\ref{sec:bdg_form}. The construction is
closely analogous to the Debye prescription in lattice dynamics, where
the continuum phonon spectrum is truncated such that the total number of
phonon modes equals the number of microscopic lattice degrees of
freedom.

The importance of this cutoff extends beyond a simple numerical
convenience. Because the low-energy continuum model neglects the
microscopic lattice structure at short distances, its ultraviolet sector
must be constrained so that the correct state counting of the underlying
Hubbard model is preserved. The present prescription achieves exactly
this. In particular, it guarantees that high-energy states contribute
with the correct total spectral weight, allowing quantities that are
sensitive to ultraviolet physics, such as the two-body binding energy,
$\mu$ and $\Delta_0$, to recover the correct Hubbard-model 
behavior~\cite{iskin19}. 
As we demonstrate throughout this work, this state-counting procedure 
allows the continuum formalism to remain quantitatively reliable well 
beyond the weak-coupling regime, while providing a controlled connection 
between momentum-space and real-space descriptions of the same 
physical system.

For instance, at zero temperature ($T=0$) and in the $F \to 0$ limit, 
the gap equation in Eq.~\eqref{eq:gap_eqn} reduces to the two-body
bound-state condition in vacuum,
$
\frac{1}{U} = \frac{1}{N}\sum_{s\bm{k}}
\frac{1}{2\varepsilon_{s\bm{k}} - E_b},
$
where $|E_b|$ with $E_b < 0$ denotes the binding energy of a spin-singlet 
pair at zero center-of-mass momentum~\cite{iskin24}. 
Evaluating this equation within the low-energy continuum model and 
introducing the same radial cutoff $k_c$, we obtain
\begin{equation}
\frac{t}{U}
=
-\frac{E_c}{8\pi E_b}
+
\frac{1}{16\pi}\ln \Bigl(1-\frac{2E_c}{E_b}\Bigr).
\label{eq:Eb_eqn}
\end{equation}
This expression reproduces the expected limiting behaviors. 
In the weak-coupling regime $|E_b| \ll E_c$, the logarithmic 
contribution is subleading, and one finds
$
E_b \approx -U/2,
$
consistent with the two-particle bound-state energy of the 
attractive Hubbard model in this regime.
In the opposite limit $|E_b| \gg E_c$, the logarithm can be expanded 
to leading order in $E_c/|E_b|$, yielding
$
E_b \approx -U,
$
which recovers the correct strong-coupling molecular limit of 
the lattice problem. The recovery of both limits is a nontrivial 
consequence of the cutoff prescription introduced above, which preserves 
the correct number of lattice degrees of freedom in the continuum description.
Throughout this work, however, we focus on the flat-band regime 
$0 \le F \le 1$ with $|E_b| \ll E_c$, where the continuum approximation 
remains quantitatively controlled in the BdG calculations.

\begin{figure}
\centering
\includegraphics[width=\linewidth]{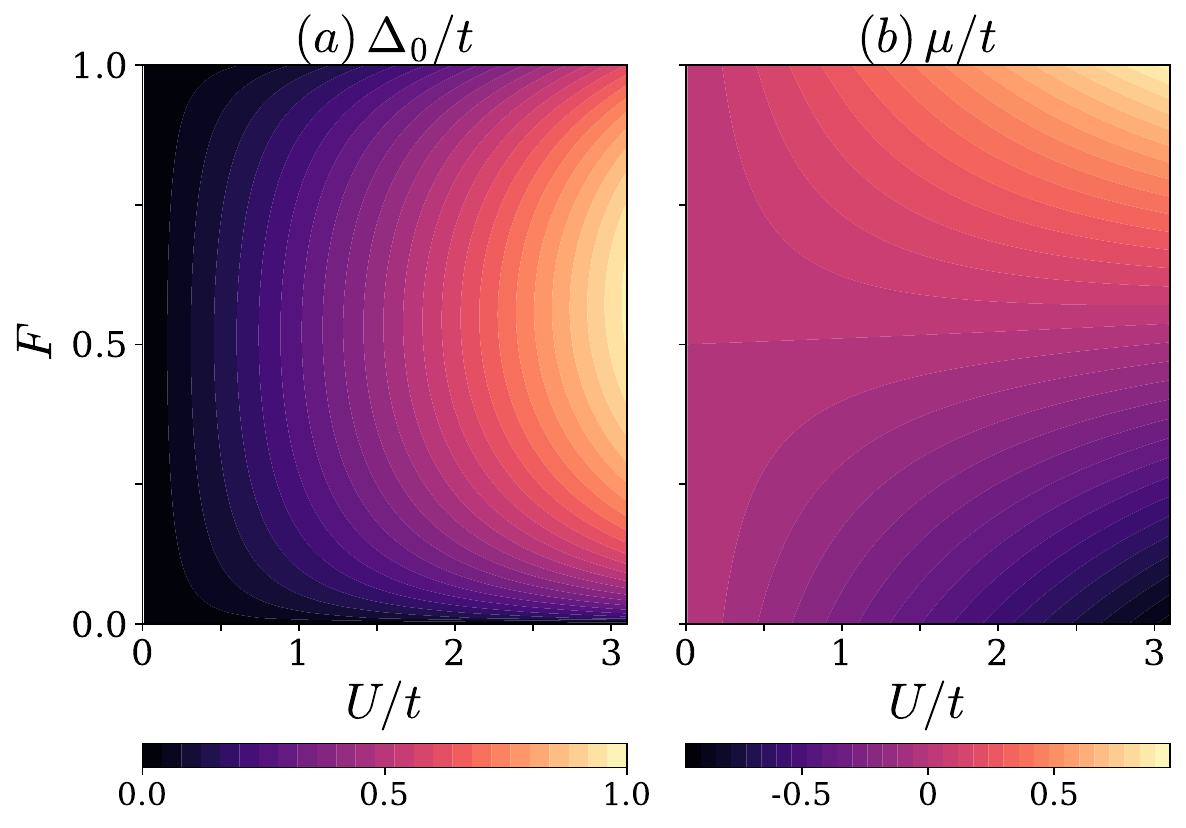}
\caption{\justifying
Self-consistent zero-temperature solutions of (a) the pairing gap
$\Delta_0$ and (b) the chemical potential $\mu$ obtained from
Eqs.~\eqref{eq:gap_closed} and~\eqref{eq:number_closed}. In the
weak-coupling limit ($U/t \to 0$), both quantities exhibit exact
particle-hole symmetry about $F=0.5$, corresponding to a half-filled
flat band. 
}
\label{fig:selfcons}
\end{figure}

At $T=0$, the thermal factor reduces to $\mathcal{X}_{s\bm{k}}=1$, and
the continuum momentum integrals in
Eqs.~\eqref{eq:gap_eqn} and~\eqref{eq:number_eqn} can be evaluated
analytically. The resulting self-consistency equations take the form
\begin{align}
\frac{1}{U} &=
\frac{1}{4E_0}
+ \frac{1}{16\pi t}
\ln \bigg[
\frac{E_c-\mu+\sqrt{(E_c-\mu)^2+\Delta_0^2}}
{-\mu+E_0}
\bigg],
\label{eq:gap_closed}
\\
F &=
\frac{1}{2} + \frac{\mu}{2E_0}
+ \frac{E_c + E_0 - \sqrt{(E_c-\mu)^2+\Delta_0^2}}
{8\pi t},
\label{eq:number_closed}
\end{align}
where $E_0 = \sqrt{\mu^2+\Delta_0^2}$.
For fixed interaction strength $U$ and filling $F$, 
Eqs.~\eqref{eq:gap_closed} and~\eqref{eq:number_closed} determine 
$\mu$ and $\Delta_0$, whose numerical solutions are shown in 
Fig.~\ref{fig:selfcons}.
As a useful reference, the noninteracting limit ($\Delta_0=0$) 
yields $F=1/2$ at $\mu=0$, reflecting the half-filling of the flat 
band, while $F = 1 + \mu/(4\pi t)$ for $\mu>0$, corresponding to 
occupation of the quadratically dispersing upper band.
Further insight is obtained from the asymptotic solutions:
in the weak-coupling regime $|E_b| \ll E_c$, we find
$
\mu=\frac{U}{2}\left(F-\frac12\right)
$
and
$
\Delta_0=\frac{U}{2}\sqrt{F(1-F)},
$
valid for $0\le F\le1$ and exhibit particle-hole symmetry about $F=1/2$. 
In the opposite regime $|E_b| \gg E_c$, where $|\mu|\gg E_c$ and $\mu<0$, 
the corresponding solutions become
$
\mu=-\frac{U}{2}(1-F)
$
and
$
\Delta_0=\frac{U}{2}\sqrt{F(2-F)},
$
valid for $0\le F\le2$ and exhibit particle-hole symmetry about $F=1$, 
corresponding to half filling of the full two-band system.
Thus, in the dilute limit $F\to0$, both asymptotic solutions reduce to
$
\mu \to E_b/2,
$
despite being obtained from opposite coupling limits. Consequently, the
many-body chemical potential is controlled entirely by the two-body bound
state, indicating that the elementary low-energy degrees of freedom are
pairs rather than individual fermions~\cite{nsr85, leggett}. 
Unlike a conventional dispersive band, where weak attraction leads to 
a BCS state built around a Fermi surface, 
the flat band lacks a kinetic-energy scale that can compete with
pair formation. As a result, the dilute flat-band superfluid possesses an
intrinsically molecular character and is most naturally viewed as a
Bose-Einstein condensate (BEC) of bound pairs, even in the weak-coupling 
limit~\cite{iskin24, iskin24c, iskin25}.

In Sec.~\ref{sec:bdg_form}, we demonstrate that the real-space BdG
calculations are in excellent agreement with the bulk self-consistent
solutions of Eqs.~\eqref{eq:gap_closed} and~\eqref{eq:number_closed}.
Before presenting those results, we first examine the zero-temperature
coherence length within the momentum-space formalism, which provides an
independent benchmark for the vortex healing length extracted from the
BdG simulations.

\subsection{Coherence length}
\label{sec:coherence}

To extract a characteristic zero-temperature length scale in the
flat-band regime, we map the dilute system of Cooper pairs onto a
weakly-interacting BEC~\cite{PethickSmith}, 
following Ref.~\cite{iskin24c, elden26}. The corresponding healing 
length is~\cite{PethickSmith}
\begin{align}
(\xi_\mathrm{B}^2)_{ij} = \frac{(M_\mathrm{B}^{-1})_{ij}}{2U_\mathrm{B} n_\mathrm{B}},
\label{eqn:xiB}
\end{align}
where $M_\mathrm{B}^{-1}$ is the inverse effective-pair-mass tensor 
(derived below), $U_\mathrm{B}$ is the effective pair-pair interaction, 
and $n_\mathrm{B}$ is the boson density. In the dilute limit, 
one may identify $U_\mathrm{B} \simeq 2U$ and 
$n_\mathrm{B} = F_\mathrm{B}/a^2$ with $F_\mathrm{B} \simeq F/2$ 
when $0\le F \le1$ and $|E_b|\ll E_c$~\cite{iskin24c, iskin24}.
Away from the strictly dilute regime, we employ the particle-hole
symmetric interpolation
$
F_\mathrm{B} = F(1-F)/2
$
for the condensed pair density, which vanishes at the band edges and 
reproduces $F_\mathrm{B} \simeq F/8$ near half filling, consistent 
with the solutions of Eqs.~\eqref{eq:gap_closed} and~\eqref{eq:number_closed}
~\footnote{
An equivalent expression for the condensed pair density is
$F_{\mathrm{B}} = F_c/2$, where the filling of condensed particles 
is given by
$
F_c = \frac{\Delta_0^2}{N}\sum_{s\mathbf{k}}
\frac{\mathcal{X}_{s\mathbf{k}}^2}{2E_{s\mathbf{k}}^2}
$
within mean-field theory~\cite{iskin18c}.
}.
Substituting the pair mass obtained below yields
\begin{equation}
\frac{\xi_\mathrm{B}}{a}
=
\frac{1}{\sqrt{4F_\mathrm{B}}}
\Bigg[
\frac{1}{8\pi}
+
\frac{1}{4\pi}
\ln \Bigl(\frac{E_c}{U}\Bigr)
\Bigg]^{1/2},
\label{eq:xi_final}
\end{equation}
which provides a simple analytical estimate for the coherence length.

Equation~\eqref{eq:xi_final} reveals that the coherence length in a
flat-band superfluid is governed primarily by filling rather than by
interaction strength. Indeed, the interaction enters only through the
logarithmic factor $\ln(E_c/U)$, implying that even substantial changes
in $U$ produce only modest variations in $\xi_\mathrm{B}$. This behavior is in
sharp contrast to conventional dispersive-band superconductors, where
the coherence length is typically much more sensitive to the pairing
strength. Instead, the dominant dependence arises through the bosonic
density factor $1/\sqrt{F_\mathrm{B}}$. As the flat band approaches either
the empty limit ($F\to0$) or the fully occupied limit ($F\to1$), the
density of mobile pairs vanishes, causing the healing length to diverge.
Physically, the condensate then becomes increasingly dilute and loses
its ability to recover from local perturbations over short distances.
Conversely, the pair density is maximal at half filling, where
$\xi_\mathrm{B}$ reaches its minimum value. This predicts that vortices and
other spatial textures should be most localized near half filling and
become progressively more extended as the flat band is emptied or
filled. As shown in Sec.~\ref{sec:bdg}, these trends are borne out by
fully self-consistent real-space BdG calculations, providing an
important consistency check on the effective bosonic description.

The pair mass $M_\mathrm{B}$ is governed by the quantum geometry of the Bloch
states. For instance, following Ref.~\cite{iskin24, iskin24c}, the exact 
inverse effective-mass tensor of the lowest two-body bound state 
decomposes as
$
(M^{-1}_{2b})_{ij} =
(M^{-1}_{2b})_{ij}^{\mathrm{intra}}
+
(M^{-1}_{2b})_{ij}^{\mathrm{inter}},
$
where the intraband contribution is controlled by the band dispersion
and the interband contribution by the quantum metric $g^{\bm{k}}_{ij}$
of Eq.~\eqref{eq:metric}:
\begin{align}
(M^{-1}_{2b})_{ij}^{\mathrm{intra}}
&=
2\frac{
\sum_{\bm{k}}
\frac{\partial_i \varepsilon_{+,\bm{k}}
\partial_j \varepsilon_{+,\bm{k}}}
{(2\varepsilon_{+,\bm{k}}-E_b)^3}
}{
\sum_{s\bm{k}}
\frac{1}{(2\varepsilon_{s\bm{k}}-E_b)^2}
},
\label{eq:pair_mass_intra}
\\[6pt]
(M^{-1}_{2b})_{ij}^{\mathrm{inter}}
&=
\frac{
\sum_{s\bm{k}}
\frac{g_{ij}^{\bm{k}}}{2\varepsilon_{s\bm{k}}-E_b}
-
\sum_{\bm{k}}
\frac{2g_{ij}^{\bm{k}}}
{\varepsilon_{+,\bm{k}}+\varepsilon_{-,\bm{k}}-E_b}
}{
\sum_{s\bm{k}}
\frac{1}{(2\varepsilon_{s\bm{k}}-E_b)^2}
}.
\label{eq:pair_mass_inter}
\end{align}
Evaluating these expressions in the weak-coupling regime relevant to
the flat-band regime, where $|E_b|\ll E_c$ and $E_b\simeq-U/2$, yields
$
(M^{-1}_{2b})_{xx}^{\mathrm{intra}}
=
\frac{Ua^2}{8\pi}
$
and
$
(M^{-1}_{2b})_{xx}^{\mathrm{inter}}
=
\frac{Ua^2}{4\pi}
\ln\big(\frac{E_c}{U}\big).
$
The logarithmic contribution originates from the quadratic band touching
structure in two dimensions, which controls the infrared behavior of the
interband processes~\cite{iskin19, wu21}.
Combining the two contributions gives the isotropic inverse pair mass
\begin{equation}
M_{2b}^{-1}
=
\frac{Ua^2}{8\pi}
+
\frac{Ua^2}{4\pi}
\ln \Big (\frac{E_c}{U} \Big).
\label{eq:pair_mass_final}
\end{equation}
Motivated by the behavior reported for the pyrochlore lattice
in Ref.~\cite{iskin24c}, where the Cooper-pair mass was found to
coincide, to very good approximation, with the two-body bound-state
mass throughout the dilute flat-band regime, we identify the pair 
mass $M_\mathrm{B}$ entering the coherence length with $M_{2b}$.

Equation~\eqref{eq:pair_mass_final} illustrates a central feature of
flat-band superconductivity: although the flat band itself carries no
conventional kinetic energy, bound pairs nevertheless acquire a finite
mobility through virtual interband processes encoded in the quantum
metric. Indeed, for $|E_b| \ll E_c$, the logarithmically enhanced
interband contribution exceeds the intraband term, implying that the
pair mass is controlled predominantly by the geometry of the Bloch
states rather than by the dispersion of the upper band.
This provides a microscopic realization of the broader principle that
quantum geometry can replace band dispersion as the source of transport
in flat-band systems.

The momentum-space estimate of Eq.~\eqref{eq:xi_final} therefore
provides a parameter-free prediction for the coherence length. In the
next section, we test this prediction directly by solving the BdG
equations in real space and extracting the vortex-core size from the
self-consistent spatial profile of the order parameter.

\section{Real-space formalism}
\label{sec:bdg_form}

The momentum-space formalism of Sec.~\ref{sec:kspace} provides an efficient
description of homogeneous bulk properties but is not suitable for
spatially varying order parameters. To resolve the real-space structure
of the superfluid and independently benchmark the bulk predictions, we
therefore employ a self-consistent BdG approach~\cite{deGennes66, gygi91, sensarma06}.
A particularly important application is a singly quantized vortex,
$
\Delta(\bm{r})=\Delta(r)e^{-i\theta},
$
whose phase winding and suppressed core explicitly break translational
invariance on the scale of the coherence length $\xi_\mathrm{B}$. Consequently,
momentum is no longer a good quantum number, and a real-space formulation
is required. Throughout this section, we solve the BdG equations 
self-consistently on a finite disk in the presence of spatially 
inhomogeneous pairing fields.

\subsection{BdG equations}
\label{sec:bdg}

We retain the on-site attractive Hubbard interaction $U>0$ and the
uniform-sublattice pairing 
$
\Delta_A = \Delta_B \equiv \Delta_0
$ 
established in Sec.~\ref{sec:mf_ham}, but now allow the order 
parameter to vary in space, $\Delta_0 \rightarrow \Delta(\bm{r})$. 
The resulting mean-field Hamiltonian is
\begin{align}
\mathcal{H}_{\mathrm{mf}} &= \int d^2\bm{r}
\sum_{\sigma SS'}
\psi_{S\sigma}^{\dagger}(\bm{r})
\left[H_{SS'}(\bm{r})-\mu\delta_{SS'}\right]
\psi_{S'\sigma}(\bm{r})
\nonumber\\
&\quad
+
\int d^2\bm{r}
\sum_S
\Big[
\Delta(\bm{r})
\psi_{S\uparrow}^{\dagger}(\bm{r})
\psi_{S\downarrow}^{\dagger}(\bm{r})
+\mathrm{H.c.}
\Big],
\label{eq:Hmf_real}
\end{align}
where $S,S'\in\{A,B\}$ label the sublattices. The single-particle
operator $H_{SS'}(\bm{r})$ follows from the continuum Bloch Hamiltonian
through the substitution $\bm{k}\rightarrow-i\bm{\nabla}$. Using
Eq.~\eqref{eq:lowE_dvec}, we obtain
\begin{equation}
H(\bm{r})
=
-ta^2
\begin{pmatrix}
\partial_x^2+\partial_y^2+2\partial_x\partial_y &
\partial_x^2-\partial_y^2
\\
\partial_x^2-\partial_y^2 &
\partial_x^2+\partial_y^2-2\partial_x\partial_y
\end{pmatrix}
\label{eq:H_real}
\end{equation}
in the sublattice basis.
This operator constitutes the real-space representation of the same
low-energy continuum theory used in Sec.~\ref{sec:kspace}, preserving its
two-sublattice structure and quantum-geometric content while allowing
for spatially inhomogeneous pairing fields.

We diagonalize $\mathcal{H}_{\mathrm{mf}}$ through the Bogoliubov
transformation~\cite{deGennes66}
\begin{align}
\psi_{S\uparrow}(\bm{r})
&=
\sum_n
\big[
u_n^S(\bm{r})\gamma_{n\uparrow}
-
v_n^{S*}(\bm{r})\gamma_{n\downarrow}^{\dagger}
\big],
\\
\psi_{S\downarrow}(\bm{r})
&=
\sum_n
\big[
u_n^S(\bm{r})\gamma_{n\downarrow}
+
v_n^{S*}(\bm{r})\gamma_{n\uparrow}^{\dagger}
\big],
\end{align}
where $\gamma_{n\sigma}$ annihilates a quasiparticle in eigenstate $n$
with spin $\sigma$, and $u_n^S(\bm{r})$ and $v_n^S(\bm{r})$ are the
corresponding particle and hole amplitudes on sublattice $S$.
Requiring the transformation to diagonalize the mean-field Hamiltonian,
$
[\mathcal{H}_{\mathrm{mf}},\gamma_{n\sigma}]
=
-E_{n\sigma}\gamma_{n\sigma}
$
leads to the BdG eigenvalue problem
\begin{widetext}
\begin{equation}
\begin{bmatrix}
H_{AA}(\bm{r})-\mu & H_{AB}(\bm{r}) & \Delta(\bm{r}) & 0 \\
H_{BA}(\bm{r}) & H_{BB}(\bm{r})-\mu & 0 & \Delta(\bm{r}) \\
\Delta^*(\bm{r}) & 0 & -H_{AA}^{*}(\bm{r})+\mu & -H_{AB}^{*}(\bm{r}) \\
0 & \Delta^*(\bm{r}) & -H_{BA}^{*}(\bm{r}) & -H_{BB}^{*}(\bm{r})+\mu
\end{bmatrix}
\begin{bmatrix}
u_n^A(\bm{r}) \\
u_n^B(\bm{r}) \\
v_n^A(\bm{r}) \\
v_n^B(\bm{r})
\end{bmatrix}
=
E_n
\begin{bmatrix}
u_n^A(\bm{r}) \\
u_n^B(\bm{r}) \\
v_n^A(\bm{r}) \\
v_n^B(\bm{r})
\end{bmatrix},
\label{eq:bdg_4x4}
\end{equation}
\end{widetext}
which acts in the combined sublattice and particle-hole space.
The BdG Hamiltonian possesses an intrinsic particle-hole symmetry:
if $(u_n^S, v_n^S)$ is a solution at energy $E_{n\uparrow}$, then
$(v_n^{S*}, -u_n^{S*})$ is a solution at energy $-E_{n\downarrow}$.
We exploit this symmetry by solving only the spin-$\uparrow$ sector,
henceforth denoting $E_{n\uparrow} \equiv E_n$ and allowing $E_n$ to
take both signs.
Positive-energy solutions ($E_n > 0$) describe spin-$\uparrow$
quasiparticles directly, while each negative-energy solution
($E_n < 0$) is identified, via the particle-hole symmetry, with
a positive-energy spin-$\downarrow$ quasiparticle at energy
$E_{n\downarrow} = -E_n > 0$.
The full quasiparticle spectrum for both spin species is therefore
contained in the single set of eigenvalues $\{E_n\}$,
with no separate diagonalization of the spin-$\downarrow$ sector required.

The pairing field must be determined self-consistently from the
quasiparticle eigenfunctions. Substituting the Bogoliubov
transformation into
$
\Delta(\bm{r})
=
U
\langle
\psi_{S\downarrow}(\bm{r})
\psi_{S\uparrow}(\bm{r})
\rangle
$
and evaluating the thermal average with the Fermi distribution
$
f(E)=1/(e^{E/T}+1)
$
yields
\begin{equation}
\Delta(\bm{r})
=
-U
\sum_n
u_n^S(\bm{r})
v_n^{S*}(\bm{r})
f(E_n),
\label{eq:gap}
\end{equation}
which is independent of the choice of sublattice owing to the uniform
on-site pairing interaction. Similarly, $\rho_{\sigma}(\bm{r}) = \sum_S \langle 
{\psi_{S\sigma}}^\dagger(\bm{r}) \psi_{S\sigma}(\bm{r}) \rangle$  
yields the local spin densities as
\begin{align}
\rho_{\uparrow}(\bm{r})
&=
\sum_{nS}
|u_n^S(\bm{r})|^2
f(E_n),
\\
\rho_{\downarrow}(\bm{r})
&=
\sum_{nS}
|v_n^S(\bm{r})|^2
\bigl[1-f(E_n)\bigr],
\label{eq:density}
\end{align}
from which the total density follows as
$
\rho(\bm{r})
=
\rho_{\uparrow}(\bm{r})
+
\rho_{\downarrow}(\bm{r}).
$
The chemical potential is adjusted such that the global filling
$
F
=
\frac{1}{N}
\int d^2\bm{r}
\rho(\bm{r})
$
matches the target value, ensuring consistency with
Eq.~\eqref{eq:number_eqn}. The coupled BdG and self-consistency
equations, Eqs.~\eqref{eq:bdg_4x4}-\eqref{eq:density}, are then solved
iteratively until both the order parameter and particle number have
converged.

\subsection{Vortex profiles}
\label{sec:vortex}

We solve the BdG eigenvalue problem of Eq.~\eqref{eq:bdg_4x4} on a
disk of radius $R = 45a$ by expanding the four-component spinor in a
Bessel-function basis adapted to the disk 
geometry~\cite{gygi91, sensarma06}, as detailed in
Appendix~\ref{app:matrix}. The basis is truncated at $M = 60$
angular-momentum channels and $J = 50$ radial modes per channel,
retaining only those modes whose single-particle energy lies below
the cutoff $E_c = 4\pi t$ of Sec.~\ref{sec:self_cons}, as expressed
by Eq.~\eqref{eq:truncation}. This is the same state-counting
prescription that fixes the momentum-space cutoff, so the real-space
and momentum-space calculations describe the same low-energy sector
and can be compared without further adjustment. The construction and
diagonalization of the BdG matrix are GPU-accelerated via the cuBLAS
and cuSOLVER libraries. The spectrum and pairing field are made
consistent with each other iteratively using the algorithm
described in Fig.~\ref{alg:bdg}.

\begin{figure}[h]
\begin{algorithmic}[1]
\State Initialize $\Delta^{(0)}(r)$ from the weak-coupling
       momentum-space solution of Sec.~\ref{sec:self_cons}.
\For{$i = 1, 2, \ldots$ until convergence}
    \State Build $\mathcal{H}_{\mathrm{mf}}[\Delta^{(i-1)}]$ in the Bessel basis.
    \State Diagonalize to obtain $\{E_n, [u_n^S(r),v_n^S(r)]\}$.
    \State Update $\Delta^{(i)}(r)$ via Eq.~(\ref{eq:gap}).
    \If{$\max_r|\Delta^{(i)}-\Delta^{(i-1)}|
         / \max_r|\Delta^{(i-1)}| < \epsilon$}
        \State \textbf{break}
    \EndIf
\EndFor
\State \Return converged $\Delta(r)$ and $\rho(r)$.
\end{algorithmic}
\caption{Algorithm for the self-consistent BdG solver.}
\label{alg:bdg}
\end{figure}
%%%
%

\begin{figure}[h]
\centering
\begin{overpic}[width=0.99\linewidth]{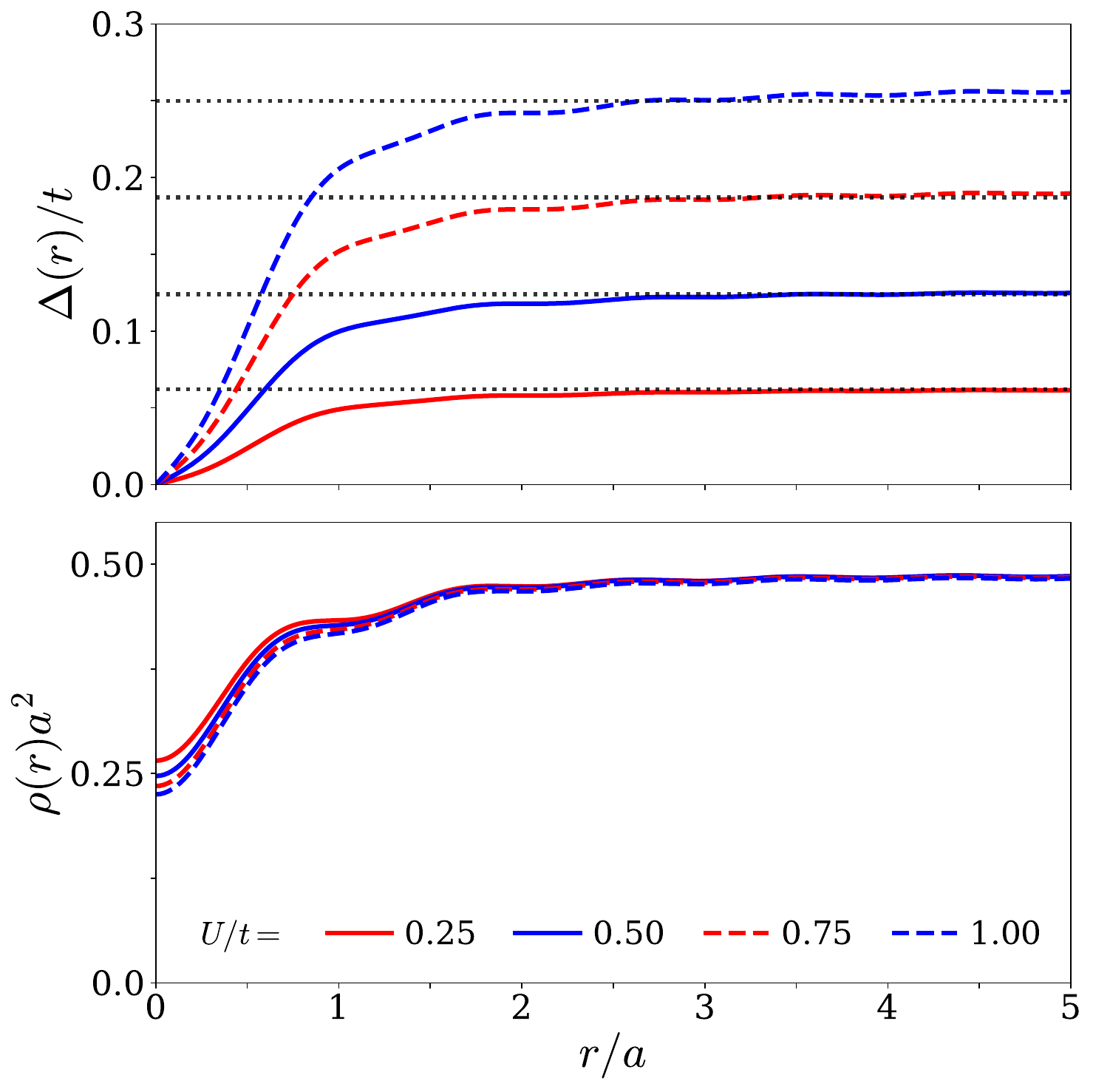}
\put(52, 18){\includegraphics[width=0.43\linewidth]{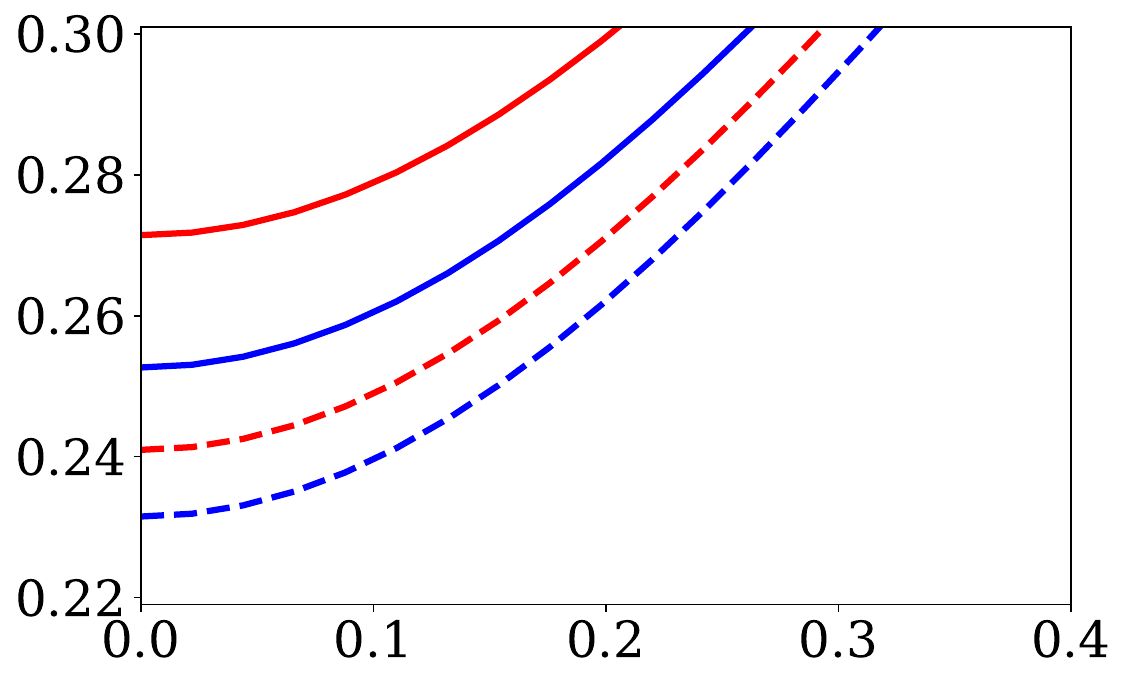}}
\end{overpic}
\caption{\justifying
(a) Radial profile of the superconducting order parameter $\Delta(r)$
for several interaction strengths $U$. The corresponding bulk
momentum-space values are indicated by dotted horizontal black lines 
for comparison. 
(b) Local particle density $\rho(r)$ for the same parameters. The inset 
shows a zoom near the vortex center ($r=0$), highlighting the finite core
density arising from localized bound states. All results are obtained
at fixed bulk filling $F=0.49$.
}
\label{fig:U_sweep}
\end{figure}

We now present the self-consistent spatial profiles of $\Delta(r)$
and $\rho(r)$. We first examine the interaction dependence at fixed
filling, followed by the filling dependence at fixed interaction
strength. Figure~\ref{fig:U_sweep}(a) shows $\Delta(r)$ for
$U/t=0.25$--$1$ at fixed $F=0.49$. In all cases, the order parameter
vanishes at the vortex center due to the imposed $2\pi$ phase winding
and recovers to a uniform bulk value within a few lattice spacings.
The asymptotic values coincide very well with the momentum-space 
solutions $\Delta_0$ obtained from Eqs.~\eqref{eq:gap_closed}
and~\eqref{eq:number_closed}, as indicated by the dotted horizontal black 
lines, providing a direct consistency check between the real-space BdG 
calculations and the bulk momentum-space theory.
The corresponding density profiles are shown in
Fig.~\ref{fig:U_sweep}(b). While the order parameter is completely
suppressed at the vortex center, the particle density remains finite
and approaches the bulk value $\rho_0=F/a^2$ away from the core. This
behavior reflects the presence of localized bound states that partially
fill the vortex core~\cite{caroli64}. The density depletion nevertheless 
becomes progressively stronger with increasing $U$~\cite{sensarma06}, 
as highlighted in the inset. 
Thus, increasing the interaction strength primarily deepens the
core depletion without qualitatively altering the vortex structure,
which remains much more pronounced in $\Delta(r)$ than in $\rho(r)$.

\begin{figure}[h]
    \centering
    \includegraphics[width=0.99\linewidth]{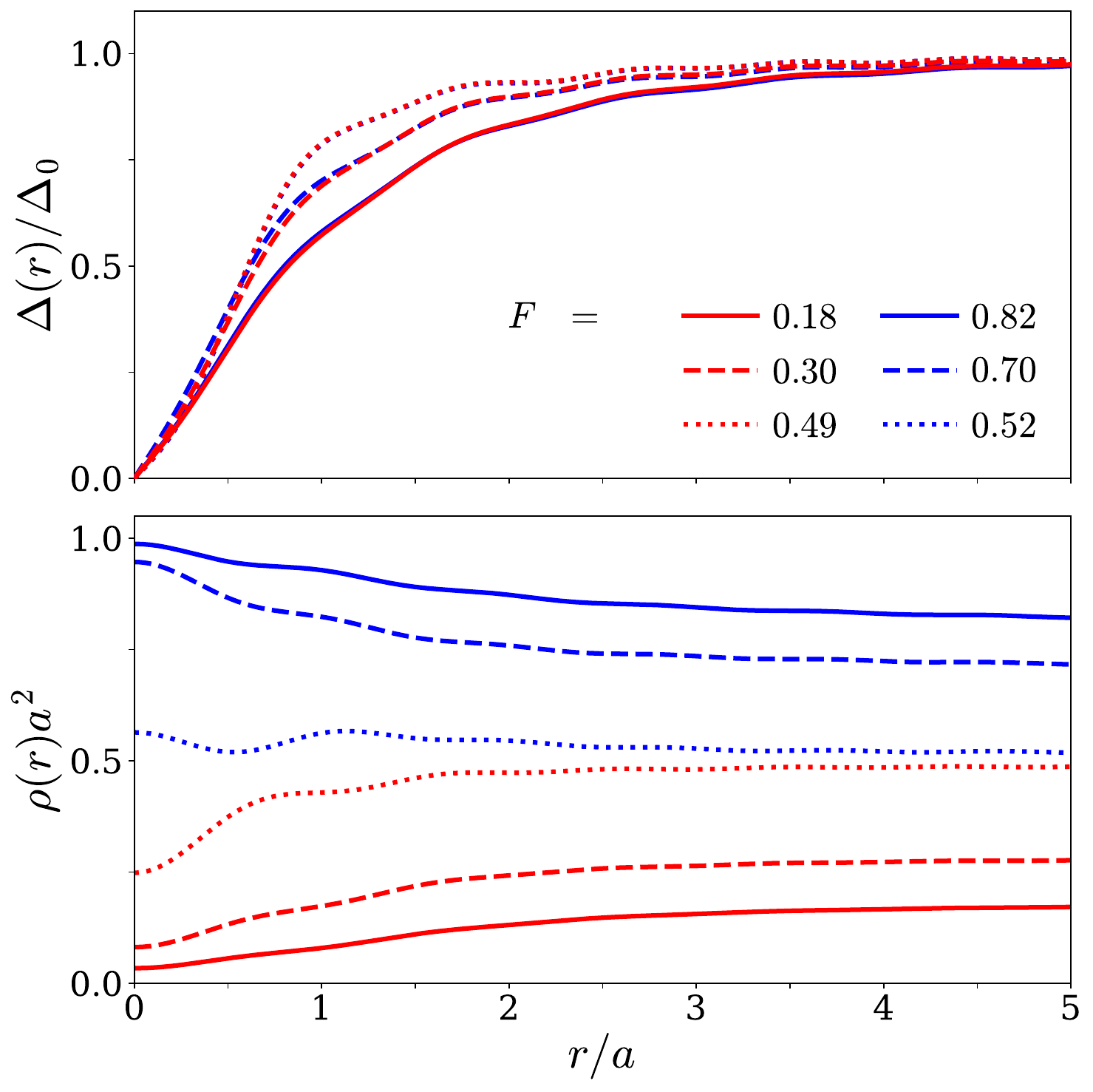}
\caption{\justifying
(a) Normalized superconducting order parameter $\Delta(r)/\Delta_0$
for different global fillings $F$, illustrating the evolution of the
vortex-core size with particle density. The fillings are chosen with respect
to the particle-hole symmetric point. (b) Corresponding local particle
density $\rho(r)a^2$ across the same filling values, computed at fixed
interaction strength $U=0.5t$.
}
\label{fig:F_sweep}
\end{figure}

The filling dependence at fixed $U/t=0.5$ is shown in
Fig.~\ref{fig:F_sweep}. Figure~\ref{fig:F_sweep}(a) shows the normalized 
profiles $\Delta(r)/\Delta_0$, which remove the trivial variation of 
the bulk order parameter $\Delta_0$ and isolate the evolution of 
the vortex-core size. The core is most localized near half filling 
and expands as $F$ approaches either band edge. 
Moreover, the profiles for $F$ and $1-F$
nearly collapse, reflecting the approximate particle-hole symmetry of
Eqs.~\eqref{eq:gap_closed} and~\eqref{eq:number_closed} about
$F=1/2$ in the weak-coupling regime.
The density profiles in Fig.~\ref{fig:F_sweep}(b) reveal the same
symmetry more directly~\cite{hayashi98, lages06}. 
For $F<1/2$, the vortex core is depleted,
with $\rho(r)$ falling below its bulk value at the origin and
recovering outward. For $F>1/2$, the behavior is reversed and the core
exhibits an excess density relative to the bulk. These two cases are
related by the particle-hole transformation
$\rho(\mathbf{r})\rightarrow a^{-2}-\rho(\mathbf{r})$ under
$F\rightarrow 1-F$, implying that the profile at filling $1-F$ is the
reflection of that at filling $F$ about the line $\rho a^2=1/2$. The
near-perfect correspondence between particle-hole conjugate fillings
shows that this symmetry remains quite robust even at $U=0.5t$.

Taken together, Figs.~\ref{fig:U_sweep} and~\ref{fig:F_sweep}
identify filling, rather than interaction strength, as the primary
control parameter for the vortex structure. The relevant density scale
is not $F$ itself but the particle-hole-symmetric combination
$F_\mathrm{B}=F(1-F)/2$, which vanishes at the band edges and is maximal at
half filling. This is precisely the combination that enters the
momentum-space coherence length in Eq.~\eqref{eq:xi_final}, providing
a direct connection between the real-space vortex-core size and the
quantum-geometric description developed in Sec.~\ref{sec:coherence}.

\subsection{Vortex-core size}
\label{sec:xi_scaling}

\begin{figure}[h]
    \centering
    \includegraphics[width=\linewidth]{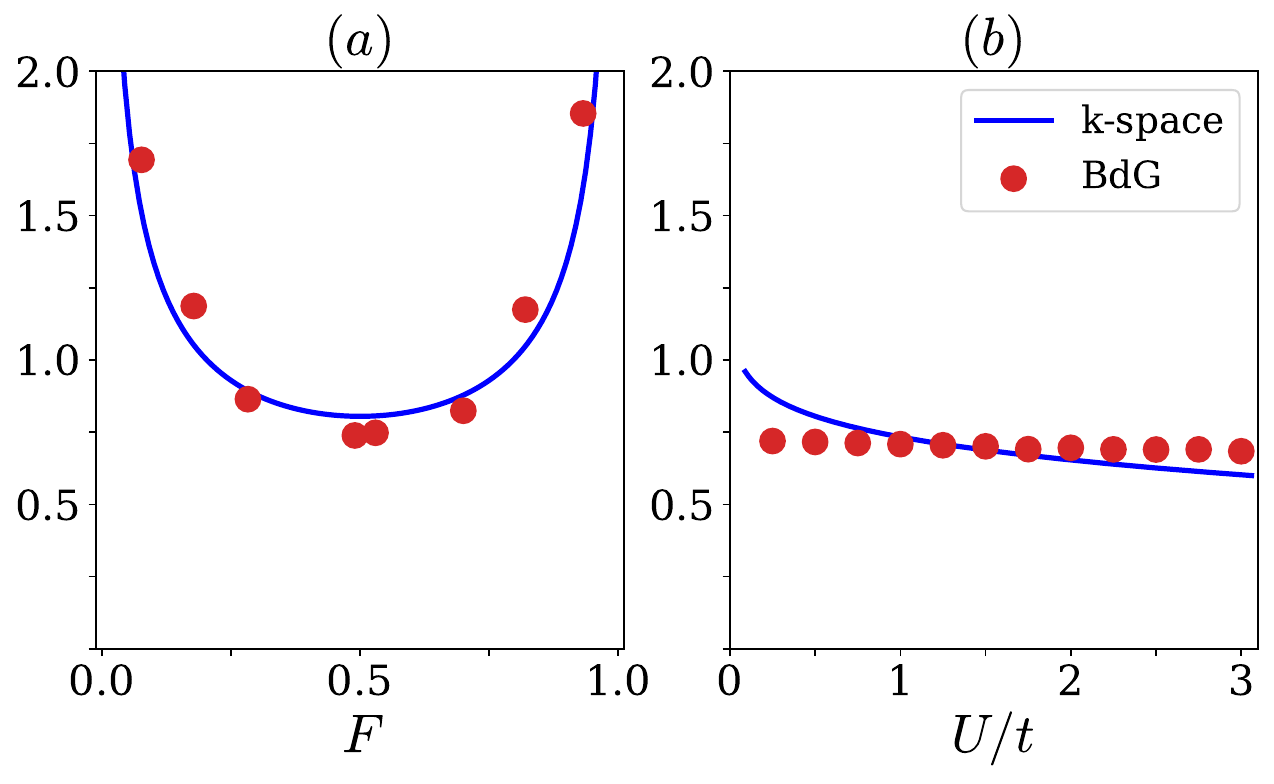}
    \caption{\justifying
        Coherence length from real-space BdG calculations ($\xi_v$, markers)
        compared with the momentum-space estimate of Eq.~\eqref{eq:xi_final}
        ($\xi_\mathrm{B}$, solid line). (a) Filling dependence at fixed $U=0.5t$: 
        $\xi_v/a$ shows a pronounced minimum at half-filling ($F=0.5$) and 
        diverges toward both band edges ($F\to 0$ or $1$), consistent with the
        $F_\mathrm{B}^{-1/2}$ scaling of Eq.~\eqref{eq:xi_final}. (b) Interaction
        dependence at fixed $F=0.49$: $\xi_v/a$ varies only weakly with
        $U/t$, reflecting the logarithmic factor in
        Eq.~\eqref{eq:xi_final}.
    }
    \label{fig:coherence_length}
\end{figure}

To compare quantitatively with the momentum-space prediction for 
$\xi_\mathrm{B}$, we extract a coherence length scale $\xi_v$ from 
each converged order-parameter profile. 
In the dilute limit, the recovery of $\Delta(r)$ is well described 
by the bosonic healing form $\Delta(r)=\Delta_0\tanh[r/(\sqrt{2}\xi_v)]$
\footnote{This form and the factor of $\sqrt{2}$ originate from the 
Gross-Pitaevskii description of a singly quantized vortex in a 
weakly-interacting BEC. Writing the condensate wavefunction as 
$\psi(\mathbf{r})=f(r)e^{-i\theta}$, the phase winding produces an 
effective centrifugal contribution proportional to $1/r^2$ in the 
radial kinetic operator, so that $f(r)$ obeys 
$
-\frac{1}{2M_\mathrm{B}}\big(\partial_r^2 f
+\frac{1}{r}\partial_r f-\frac{f}{r^2}\big) 
+ U_\mathrm{B} f^3 = \mu_\mathrm{B} f,
$ 
with $\mu_\mathrm{B}=U_\mathrm{B} n_\mathrm{B}$ in the bulk. Although 
this equation has no exact closed-form solution, the interpolating 
form 
$
f(r) \approx \sqrt{n_\mathrm{B}} 
\tanh[r/(\sqrt{2}\xi_\mathrm{B})]
$ 
correctly captures both the linear vanishing of $f(r)$ at the vortex 
core imposed by the centrifugal term and the asymptotic recovery to 
the bulk value $\sqrt{n_\mathrm{B}}$ over the healing length 
$
\xi_\mathrm{B}=1/\sqrt{2M_\mathrm{B}U_\mathrm{B}n_\mathrm{B}}
$ 
of Eq.~\eqref{eqn:xiB}. For example, see page 176 in 
Ref.~\cite{PethickSmith}. In the main text, $\Delta(r)$ plays the 
role of $f(r)$ and $\Delta_0$ plays the role of 
$\sqrt{n_\mathrm{B}}$.
}
See also Ref.~\cite{simonucci13}.
Away from this limit, however, the vortex profile evolves with filling, 
making a fit to a fixed functional form unreliable. We therefore adopt a 
fit-free definition consistent with the same convention: $\xi_v$ is the 
radius at which
\begin{equation}
    \Delta(\xi_v) = \Delta_0\tanh\bigg(\frac{1}{\sqrt{2}}\bigg) \approx 0.61\Delta_0,
    \label{eq:xi_def}
\end{equation}
i.e., the value attained at $r=\xi_v$ by the bosonic healing profile.
This prescription recovers the conventional healing length whenever the
$\tanh$ form applies and remains well defined for arbitrary profile
shapes, so the comparison between the BdG results and the
momentum-space prediction of Eq.~\eqref{eq:xi_final} in
Fig.~\ref{fig:coherence_length} introduces no additional fitting
parameters.

Figure~\ref{fig:coherence_length}(a) shows $\xi_v/a$ as a function of
filling at fixed $U=0.5t$. Two features predicted by the $1/\sqrt{F_\mathrm{B}}$
scaling of Eq.~\eqref{eq:xi_final} are clearly visible: a symmetric
minimum at half-filling, where $\xi_v$ falls below one lattice constant,
and a divergence toward both band edges as the effective pair density
vanishes. The BdG results follow the momentum-space prediction
throughout the flat band, including the particle-hole symmetry about
$F=1/2$ already evident in the collapsed profiles of
Fig.~\ref{fig:F_sweep}(a). This agreement is parameter-free: the
analytical curve is fixed entirely by $U$, $F$, and the cutoff
$E_c=4\pi t$, with the filling dependence entering only through the
particle-hole-symmetric combination $F_\mathrm{B}=F(1-F)/2$.

Figure~\ref{fig:coherence_length}(b) shows $\xi_v/a$ as a function of
$U/t$ at fixed $F=0.49$. Both descriptions yield a weak dependence, as
anticipated from the logarithmic factor in Eq.~\eqref{eq:xi_final}: the
BdG coherence length remains close to $0.7a$ over more than a decade in
coupling strength, while the momentum-space estimate 
decreases only slowly with increasing $U$. See also Ref.~\cite{li25}.
Quantitative agreement is less precise here; 
the analytical curve overestimates the core size at 
weak coupling and intersects the BdG data near $U\sim t$. The residual 
discrepancy is attributable to finite-size effects in the BdG calculation. 
The logarithmic factor in Eq.~\eqref{eq:xi_final} originates from the
quadratic band touching, whose infrared contribution is cut off only at
vanishing energy in the thermodynamic limit. On the finite disk,
however, the single-particle spectrum is discrete and the lowest level
lies a finite distance from the band-touching point, introducing an
effective infrared cutoff. This suppresses the asymptotic logarithmic
growth of the pair mass, and hence of $\xi_v$, most strongly at weak
coupling, where the continuum prediction is most sensitive to 
low-energy states. 
The filling dependence, by contrast, is governed primarily by
the particle-hole-symmetric factor $F_\mathrm{B}$ and is therefore much less
affected by the discrete spectrum.

The two panels of Fig.~\ref{fig:coherence_length} together identify the
physical origin of the vortex-core size in this flat-band
superconductor. The standard BCS estimate $\xi_v \sim v_{\mathrm{F}}/\Delta_0$
~\cite{tinkham04, leggett} breaks down because an isolated flat 
band lacks a well-defined Fermi
surface and therefore the conventional intraband kinetic processes that
govern Cooper pair transport. Instead, the finite coherence length is set
by the pair mass, whose dominant contribution arises from virtual interband
processes encoded in the quantum metric of the Bloch states,
Eq.~\eqref{eq:metric}. The vortex core therefore provides a direct
real-space signature of quantum geometry: its size is governed by the
geometric pair mass, its weak logarithmic sensitivity to $U$ reflects
the logarithmic growth of that mass, and its strong filling dependence
tracks the particle-hole-symmetric pair density $F_\mathrm{B}$.

\section{Conclusion}
\label{sec:conclusion}

We have studied the real-space structure of a singly quantized vortex
in a flat-band superconductor described by the attractive Hubbard model
on the Mielke checkerboard lattice. Using a self-consistent
BdG formulation on a finite disk, we computed the
spatial profiles of the order parameter and particle density across a
wide range of fillings and interaction strengths, and extracted a
coherence length $\xi_v$ from each converged profile using a fit-free
prescription based on the bosonic healing form.
We compared $\xi_v$ against a closed-form momentum-space estimate
$\xi_\mathrm{B}$ derived by mapping the dilute flat-band superfluid onto 
a weakly-interacting BEC, with the pair mass determined by the quantum
metric of the Bloch states and no adjustable parameters. The two
estimates agree well throughout the flat-band regime. Both independently
exhibit a pronounced minimum at half-filling, diverge toward the band
edges as the particle-hole-symmetric pair density $F_\mathrm{B} = F(1-F)/2$ 
vanishes, and depend on interaction strength only logarithmically. 
The filling dependence is well captured by the $1/\sqrt{F_\mathrm{B}}$ 
scaling of the momentum-space prediction; the residual discrepancy 
at weak coupling is consistent with finite-size effects in the BdG 
calculation, which introduce an effective infrared cutoff absent 
in the bulk theory.

These results identify quantum geometry as the physical origin of the
vortex-core size in this system. Because an isolated flat band lacks a
well-defined Fermi surface, the standard BCS estimate
$\xi_v \sim v_{\mathrm{F}}/\Delta_0$ does not apply. Instead, $\xi_v$ is
set by the geometric pair mass, whose dominant contribution arises from
virtual interband transitions encoded in the quantum metric. The vortex
core therefore constitutes a direct real-space probe of flat-band quantum
geometry: its size tracks the pair density through $F_\mathrm{B}$, while its weak
sensitivity to the pairing strength reflects the logarithmic growth of the
geometric pair mass.
Several directions are worth pursuing. It would be interesting to test
whether this geometric mechanism persists in lattices with richer band
structures or in systems where the flat band is separated from dispersive
bands by a finite gap. In addition, a systematic finite-size scaling study
of the weak-coupling discrepancy between $\xi_v$ and $\xi_\mathrm{B}$ could help
clarify the interplay between the infrared structure associated with the
quadratic band touching and the discrete spectrum of a finite system.

\begin{acknowledgments}
We acknowledge support from the U.S. Air Force Office of Scientific
Research (AFOSR) under Grant No.~FA8655-24-1-7391.
\end{acknowledgments}

\appendix 

\section{Bessel-function expansion}
\label{app:matrix}

In this appendix we derive the Bessel-function basis representation of
the BdG Hamiltonian on a finite disk~\cite{gygi91, sensarma06}. 
Since the vortex solutions considered in the main text possess 
cylindrical symmetry, it is natural to work in polar coordinates 
and expand the quasiparticle wavefunctions
in a basis of Bessel functions with definite angular momentum. We
therefore begin by rewriting the continuum single-particle Hamiltonian
of Eq.~\eqref{eq:H_real} in polar form.

Introducing the radial-angular differential operators
\begin{equation}
\hat{P}
=
\partial_r^2
-
\frac{1}{r}\partial_r
-
\frac{1}{r^2}\partial_\theta^2,
\qquad
\hat{Q}
=
\frac{2}{r}\partial_r\partial_\theta
-
\frac{2}{r^2}\partial_\theta,
\label{eq:RShat}
\end{equation}
the Cartesian derivatives can be expressed as
$\nabla^2 = \partial_r^2 + \frac{1}{r}\partial_r + \frac{1}{r^2}\partial_\theta^2$
for the Laplacian,
$\partial_x^2-\partial_y^2 = \cos(2\theta)\hat{P} - \sin(2\theta)\hat{Q}$
in the off-diagonal terms, and
$2\partial_x\partial_y = \sin(2\theta)\hat{P} + \cos(2\theta)\hat{Q}$
in the diagonal terms. Substituting these identities into
Eq.~\eqref{eq:H_real} yields
\begin{align}
\label{eq:polar_ham}
H(\bm{r})
=
-ta^2
\begin{pmatrix}
\nabla^2 + C_\theta \hat{Q} + S_\theta \hat{P} &
C_\theta \hat{P} - S_\theta \hat{Q} \\
C_\theta \hat{P} - S_\theta \hat{Q} &
\nabla^2 - C_\theta \hat{Q} - S_\theta \hat{P}
\end{pmatrix},
\end{align}
where $C_\theta = \cos(2\theta)$ and $S_\theta = \sin(2\theta)$. In
this form the angular dependence is fully explicit: the Laplacian is
rotationally invariant and diagonal in angular momentum, while the
$\cos(2\theta)$ and $\sin(2\theta)$ factors carry angular momentum
$\pm 2$, so the single-particle Hamiltonian couples only channels whose
angular-momentum quantum numbers differ by two units. This selection
rule is a direct manifestation of the quadratic band-touching structure
of the continuum model and leads to a sparse block structure in the BdG
matrix. As shown below, it allows the BdG differential equations to be
projected efficiently onto a Bessel-function basis, reducing the problem
to a finite-dimensional matrix eigenvalue equation.

We solve the BdG problem on a disk of radius $R$ with Dirichlet
boundary conditions, requiring all quasiparticle amplitudes to vanish
at the edge of the system. A convenient orthonormal basis is provided
by the disk eigenfunctions of the Laplacian,
$
\langle \bm{r}|jm\rangle
=
\frac{e^{im\theta}}{\sqrt{2\pi}}\Phi_{jm}(r),
$
labeled by angular momentum $m$ and radial index $j$, where
\begin{align}
\Phi_{jm}(r)
=
\frac{\sqrt{2}}{RJ_{m+1}(\beta_{jm})}
J_m \left(\beta_{jm}\frac{r}{R}\right),
\label{eq:basis}
\end{align}
Here, $J_m$ is the Bessel function of the first kind and $\beta_{jm}$
denotes its $j$th positive zero. The angular and radial factors satisfy
the orthonormality relations
\begin{align}
\int_0^{2\pi} \frac{d\theta}{2\pi}e^{i(m'-m)\theta}
&= \delta_{mm'},
\label{eq:angular_orthogonality}\\
\int_0^R rdr\Phi_{jm}(r)\Phi_{j'm}(r)
&= \delta_{jj'},
\label{eq:radial_orthogonality}
\end{align}
together with the boundary condition $\Phi_{jm}(R)=0$ and the
Laplacian eigenvalue equation
\begin{equation}
\nabla^2
\bigl[\Phi_{jm}(r)e^{im\theta}\bigr]
=
-\Big(\frac{\beta_{jm}}{R}\Big)^2
\Phi_{jm}(r)e^{im\theta}.
\label{eq:laplacian_eigen}
\end{equation}
The basis therefore diagonalizes the rotationally invariant part of the
single-particle Hamiltonian and is particularly well suited to the
cylindrical geometry of the vortex problem. Moreover, the nontrivial
angular dependence of Eq.~\eqref{eq:polar_ham} couples only
angular-momentum channels differing by two units, so the resulting
matrix representation remains sparse.

In practice the basis must be truncated. We retain angular-momentum
channels $m\in[-M,M]$ and radial quantum numbers $j\in[1,J]$, keeping
only those basis states whose single-particle energies lie below the
continuum cutoff $E_c=4\pi t$ derived in
Sec.~\ref{sec:self_cons}. Using the Laplacian eigenvalue
$\beta_{jm}^2/R^2$, this requirement becomes
\begin{equation}
\beta_{jm}^2
\leq
4\pi\Big(\frac{R}{a}\Big)^2.
\label{eq:truncation}
\end{equation}
This truncation ensures that the real-space BdG calculation contains the
same number of low-energy degrees of freedom as the continuum
momentum-space theory. In this study we use $M=60$, $J=50$, and
$R/a=45$. For these parameters the highest retained radial mode in the
$m=0$ sector satisfies $\beta_{J0}^2 \approx 4\pi(R/a)^2$, indicating
that the basis nearly saturates the physical cutoff.

\section{BdG Hamiltonian}
\label{app:bdg_matrix}

Using the Bessel basis introduced in Appendix~\ref{app:matrix}, 
we now construct the explicit matrix representation of the BdG
Hamiltonian~\eqref{eq:bdg_4x4}. The particle and hole amplitudes,
organized as sublattice spinors
$\bm{u}_n = (u^A_n, u^B_n)^\mathrm{T}$ and
$\bm{v}_n = (v^A_n, v^B_n)^\mathrm{T}$, are expanded in the basis of
Eq.~\eqref{eq:basis} as
\begin{equation}
\begin{aligned}
\bm{u}_n(\bm{r}) &= \sum_{m=-M}^{M}\sum_{j=1}^{J}
\bm{u}_{njm}\frac{e^{im\theta}}{\sqrt{2\pi}}\Phi_{jm}(r), \\
\bm{v}_n(\bm{r}) &= \sum_{m=-M}^{M}\sum_{j=1}^{J}
\bm{v}_{njm}\frac{e^{im\theta}}{\sqrt{2\pi}}\Phi_{jm}(r),
\end{aligned}
\label{eq:bessel_expansion_vector}
\end{equation}
with two-component sublattice coefficients
$\bm{u}_{njm} = (u^A_{njm}, u^B_{njm})^\mathrm{T}$ and
$\bm{v}_{njm} = (v^A_{njm}, v^B_{njm})^\mathrm{T}$.

At fixed angular momentum $m$, the radial and sublattice components are
assembled into $2J$-dimensional vectors,
\begin{equation}
\bm{u}_{nm} =
\begin{pmatrix}
\bm{u}_{n1m}\\ \bm{u}_{n2m}\\ \vdots\\ \bm{u}_{nJm}
\end{pmatrix},
\qquad
\bm{v}_{nm} =
\begin{pmatrix}
\bm{v}_{n1m}\\ \bm{v}_{n2m}\\ \vdots\\ \bm{v}_{nJm}
\end{pmatrix}.
\label{eq:fixed_m_vectors}
\end{equation}
Collecting all $m$-blocks yields the full particle and hole vectors
\begin{equation}
\bm{U}_n =
\begin{pmatrix}
\bm{u}_{n,-M}\\ \vdots\\ \bm{u}_{n,M}
\end{pmatrix},
\qquad
\bm{V}_n =
\begin{pmatrix}
\bm{v}_{n,-M}\\ \vdots\\ \bm{v}_{n,M}
\end{pmatrix},
\label{eq:particle_hole_vectors}
\end{equation}
together with the full BdG spinor
$\bm{\Psi}_n = (\bm{U}_n, \bm{V}_n)^\mathrm{T}$.
With this ordering, the BdG eigenvalue problem
$\mathbb{H}_{\mathrm{BdG}}\bm{\Psi}_n = E_n\bm{\Psi}_n$ takes the
compact form
\begin{equation}
\mathbb{H}_{\mathrm{BdG}} =
\begin{pmatrix}
\bm{H}-\mu\bm{I} & \bm{\Delta}\\
\bm{\Delta}^{\dagger} & -\bm{H}^*+\mu\bm{I}
\end{pmatrix},
\label{eq:bdg_full_compact}
\end{equation}
where $\bm{I}$ is the identity in the combined radial,
angular-momentum, and sublattice space. The projected single-particle
and pairing matrices are indexed by $m$-blocks,
\begin{align}
\bm{H} &=
\begin{pmatrix}
\bm{h}_{-M,-M} & \bm{h}_{-M,-M+1} & \cdots & \bm{h}_{-M,M}\\
\bm{h}_{-M+1,-M} & \bm{h}_{-M+1,-M+1} & \cdots & \bm{h}_{-M+1,M}\\
\vdots & \vdots & \ddots & \vdots\\
\bm{h}_{M,-M} & \bm{h}_{M,-M+1} & \cdots & \bm{h}_{M,M}
\end{pmatrix},
\label{eq:H_block_matrix}
\\
\bm{\Delta} &=
\begin{pmatrix}
\bm{\Delta}_{-M,-M} & \bm{\Delta}_{-M,-M+1} & \cdots & \bm{\Delta}_{-M,M}\\
\bm{\Delta}_{-M+1,-M} & \bm{\Delta}_{-M+1,-M+1} & \cdots & \bm{\Delta}_{-M+1,M}\\
\vdots & \vdots & \ddots & \vdots\\
\bm{\Delta}_{M,-M} & \bm{\Delta}_{M,-M+1} & \cdots & \bm{\Delta}_{M,M}
\end{pmatrix}.
\label{eq:Delta_block_matrix}
\end{align}
Each block $\bm{h}_{mm'}$ and $\bm{\Delta}_{mm'}$ is a $2J\times 2J$
matrix carrying both radial ($jj'$) and sublattice ($SS'$) indices, so
that the total dimension of $\mathbb{H}_{\mathrm{BdG}}$ is
$4(2M+1)J$, where the factor of $4$ reflects the product of the 
particle-hole and sublattice degrees of freedom.

The block structure of Eqs.~\eqref{eq:H_block_matrix}-\eqref{eq:Delta_block_matrix}  
is sharply constrained by the angular selection rule of
Eq.~\eqref{eq:angular_orthogonality}. The single-particle Hamiltonian
Eq.~\eqref{eq:polar_ham} contains only the harmonics
$\{1,e^{\pm 2i\theta}\}$, restricting its matrix elements to
$m'-m\in\{0,\pm 2\}$, while the singly quantized vortex pairing
$\Delta(\bm{r})=\Delta(r)e^{-i\theta}$ carries the single harmonic
$e^{-i\theta}$, restricting its matrix elements to $m'=m+1$. Hence
\begin{align}
\bm{h}_{mm'} &= \bm{0}
\quad \text{unless} \quad m' = m \;\text{or}\; m \pm 2,
\label{eq:h_selection}\\
\bm{\Delta}_{mm'} &= \bm{0}
\quad \text{unless} \quad m' = m+1,
\label{eq:delta_selection}
\end{align}
so that $\bm{H}$ is block-tridiagonal in $m$ (with even and odd $m$
decoupled) and $\bm{\Delta}$ has a single upper diagonal. These
sparsity constraints reduce the number of nonzero blocks per row from
$\mathcal{O}(M)$ to $\mathcal{O}(1)$, making the matrix construction
and the matrix-vector products required during iterative diagonalization
significantly cheaper than a dense treatment would be.

\section{Single-particle Hamiltonian}
\label{app:sp_elements}

We now evaluate the nonzero blocks of $\bm{H}$ identified by the
selection rule~\eqref{eq:h_selection}, working with the polar-form
Hamiltonian of Eq.~\eqref{eq:polar_ham} and the Bessel basis of
Eq.~\eqref{eq:basis}.

\paragraph{Diagonal block ($m'=m$).}
Only the $\nabla^2$ term in Eq.~\eqref{eq:polar_ham} survives the
angular integration, and by Eq.~\eqref{eq:laplacian_eigen} it is
diagonal in both the radial and sublattice indices,
\begin{equation}
\bm{h}_{mm}^{jj'}
= ta^2\frac{\beta_{jm}^2}{R^2} \tau_0 \delta_{jj'},
\label{eq:h_diag}
\end{equation}
where $\tau_0$ is the $2\times 2$ identity matrix in sublattice
space.

\paragraph{Off-diagonal block ($m'=m+2$).}
The off-diagonal blocks of $\bm{H}$ are determined by the matrix
elements of the two Cartesian operators $2\partial_x\partial_y$ and
$\partial_x^2-\partial_y^2$ that appear alongside $\nabla^2$ in
Eq.~\eqref{eq:polar_ham}. Using the operators $\hat{P}$ and $\hat{Q}$
defined in Eq.~\eqref{eq:RShat}, we introduce
\begin{align}
A_{m,m+2}^{jj'} &\equiv
\langle jm| C_\theta \hat{Q} + S_\theta \hat{P} |j',m+2\rangle,
\label{eq:A_def}\\
B_{m,m+2}^{jj'} &\equiv
\langle jm| C_\theta \hat{P} - S_\theta \hat{Q} |j',m+2\rangle.
\label{eq:B_def_general}
\end{align}
The angular integrals needed to evaluate these matrix elements are
\begin{equation}
\int_0^{2\pi} \frac{d\theta}{2\pi}e^{2i\theta}\sin(2\theta)
= \frac{i}{2},
\quad
\int_0^{2\pi} \frac{d\theta}{2\pi}e^{2i\theta}\cos(2\theta)
= \frac{1}{2},
\label{eq:angular_integrals}
\end{equation}
both of which follow from Eq.~\eqref{eq:angular_orthogonality}. Acting
on $\Phi_{j',m+2}(r)e^{i(m+2)\theta}$, the operators of
Eq.~\eqref{eq:RShat} reduce to their radial counterparts,
\begin{align}
\hat{P}\;&\longrightarrow\;
\hat{P}_{m+2} \equiv \partial_r^2 - \tfrac{1}{r}\partial_r
+ \tfrac{(m+2)^2}{r^2},
\label{eq:Rhat_m}\\
\hat{Q}\;&\longrightarrow\;
i\hat{Q}_{m+2} \equiv
i\left[\tfrac{2(m+2)}{r}\partial_r
- \tfrac{2(m+2)}{r^2}\right],
\label{eq:Shat_m}
\end{align}
where the factor of $i$ in the second line arises from
$\partial_\theta\to i(m+2)$ acting on $e^{i(m+2)\theta}$. A direct
evaluation of the angular projections in
Eqs.~\eqref{eq:A_def}-\eqref{eq:B_def_general} using
Eq.~\eqref{eq:angular_integrals} then shows that
$A_{m,m+2}^{jj'} = iB_{m,m+2}^{jj'}$, where $B$ is real and given
by
\begin{equation}
B_{m,m+2}^{jj'}
= \tfrac{1}{2} \int_0^R  rdr
\Phi_{jm}(r)
\bigl(\hat{P}_{m+2} + \hat{Q}_{m+2}\bigr)
\Phi_{j',m+2}(r).
\label{eq:B_integral_def}
\end{equation}
Thus, from here on we work with $B$ alone. Substituting this relation 
into the off-diagonal entries of Eq.~\eqref{eq:polar_ham}
gives the off-diagonal block in the compact form
\begin{equation}
\bm{h}_{m,m+2}^{jj'}
= -ta^2B_{m,m+2}^{jj'}(\tau_x + i\tau_z),
\label{eq:h_off_diag}
\end{equation}
where the sublattice matrix $\tau_x + i\tau_z$ mirrors the off-diagonal
structure of the original Bloch Hamiltonian.

It remains to reduce the radial integral in
Eq.~\eqref{eq:B_integral_def} to a closed form. We use the two
standard Bessel recurrences,
\begin{align}
J_\nu'(x) &= \frac{\nu}{x}J_\nu(x) - J_{\nu+1}(x),
\label{eq:Bessel_derivative}\\
\frac{2\nu}{x}J_\nu(x) &= J_{\nu-1}(x) + J_{\nu+1}(x),
\label{eq:Bessel_index_recursion}
\end{align}
together with the Laplacian eigenrelation~\eqref{eq:laplacian_eigen},
where
$\nabla^2_{m+2} = \partial_r^2+\tfrac{1}{r}\partial_r
-\tfrac{(m+2)^2}{r^2}$
is the radial Laplacian in the $(m+2)$ channel. Adding and subtracting
$\tfrac{1}{r}\partial_r$ and $\tfrac{(m+2)^2}{r^2}$ inside the
differential bracket isolates the radial Laplacian, giving
\begin{align}
B_{m,m+2}^{jj'} = \frac{1}{2}\int_0^R   &rdr
\Phi_{jm}(r)\Big[\nabla^2_{m+2} + \frac{2(m+1)}{r}\partial_r 
\nonumber \\
+ &\frac{2(m+1)(m+2)}{r^2}\Big]\Phi_{j',m+2}(r).
\label{eq:B_laplacian_form}
\end{align}
The eigenrelation replaces $\nabla^2_{m+2}$ by
$-(\beta_{j',m+2}/R)^2$, while the derivative
recurrence~\eqref{eq:Bessel_derivative} at $\nu=m+2$ gives
\begin{align}
\label{eq:dPhi}
\partial_r\Phi_{j',m+2}(r) &= \frac{m+2}{r}\Phi_{j',m+2}(r)
\\
&- \frac{\beta_{j',m+2}}{R}N_{j',m+2} J_{m+3} \bigl(\beta_{j',m+2}\tfrac{r}{R}\bigr),
\nonumber 
\end{align}
with $N_{j',m+2}=\sqrt{2}/[RJ_{m+3}(\beta_{j',m+2})]$. One can
verify that the $J_{m+3}$ term generated here cancels exactly against
the one produced when the $1/r^2$ term is re-expressed through the
index recurrence~\eqref{eq:Bessel_index_recursion}, so that only a
$J_{m+1}$ contribution survives. Collecting what remains,
\begin{align}
\label{eq:B_two_integrals}
&B_{m,m+2}^{jj'} =\; -\frac{\beta_{j',m+2}^2}{2R^2}
\int_0^R   rdr  \Phi_{jm}(r)\Phi_{j',m+2}(r) \\
&+ \frac{\sqrt{2}(m+1)\beta_{j',m+2}}{R^2J_{m+3}(\beta_{j',m+2})} 
\int_0^R  dr  \Phi_{jm}(r)J_{m+1}\bigl(\beta_{j',m+2}\tfrac{r}{R}\bigr).
\nonumber
\end{align}
Inserting the explicit forms of $\Phi_{jm}$, rescaling to $x=r/R$, and
applying the index recurrence~\eqref{eq:Bessel_index_recursion} once
more to $J_{m+2}$ cancels the intermediate $J_{m+1}$ pieces between
the two integrals and collapses them to a single Bessel integral of
equal order,
\begin{align}
B_{m,m+2}^{jj'}
&= \frac{\beta_{j',m+2}^2}
{R^2J_{m+1}(\beta_{jm})J_{m+3}(\beta_{j',m+2})}
\nonumber\\
&\quad\times
\int_0^1   xJ_m(\beta_{jm}x)J_m(\beta_{j',m+2}x)dx.
\label{eq:B_integral_simplified}
\end{align}
The remaining integral is a Lommel integral: for arguments
$a\equiv\beta_{jm}$ not equal to $b\equiv\beta_{j',m+2}$ it reads
\begin{equation}
\int_0^1 xJ_m(ax)J_m(bx)dx
=
\tfrac{bJ_m(a)J_{m+1}(b)-aJ_{m+1}(a)J_m(b)}
{b^2-a^2}.
\label{eq:gradshteyn}
\end{equation}
where the first term in the numerator vanishes because $a=\beta_{jm}$
is a zero of $J_m$. For completeness, we also note the self-integral
$
\int_0^1 xJ_m^2(ax) dx = \frac12 J_{m+1}^2(a),
$
which is needed only in the exceptional case discussed below. Hence
\begin{equation}
B_{m,m+2}^{jj'} =
\frac{b^2}{R^2J_{m+1}(a)J_{m+3}(b)}
\begin{cases}
\frac{aJ_{m+1}(a)J_m(b)}{a^2-b^2}, & a\neq b,\\[6pt]
\frac{J_{m+1}^2(a)}{2}, & a=b.
\end{cases}
\label{eq:B_piecewise}
\end{equation}
Applying $J_{m+3}(b)=-J_{m+1}(b)$, which follow from Eq.~\eqref{eq:Bessel_index_recursion} evaluated at the respective zeros, the $a\neq b$ case simplifies to
\begin{equation}
B_{m,m+2}^{jj'} =
\frac{\beta_{jm}\beta_{j',m+2}^2J_m(\beta_{j',m+2})}
{R^2J_{m+1}(\beta_{j',m+2})
(\beta_{j',m+2}^2-\beta_{jm}^2)}.
\label{eq:B_compact}
\end{equation}
For $m\neq -1$, the zeros of $J_m$ and $J_{m+2}$ are distinct, so
Eq.~\eqref{eq:B_compact} is valid for all $j$ and $j'$. The only
exceptional case is $m=-1$, for which $J_{-1}(x)=-J_1(x)$ and
$\beta_{j,-1}=\beta_{j,1}$. For $j=j'$, one has $a=b$, and the matrix
element must therefore be evaluated from the second line of
Eq.~\eqref{eq:B_piecewise}, yielding
$
B_{-1,1}^{jj} = -\frac{\beta_{j 1}^{2}}{2R^2}.
$
The reverse coupling $\bm{h}_{m+2,m}$ follows from the same reduction
with the two channels interchanged,
$(m,j)\leftrightarrow(m+2,j')$. Hermiticity of $\bm{H}$ then follows
directly, with $h_{m+2,m}^{j'j}=(h_{m,m+2}^{jj'})^*$ verified
analytically from Eq.~\eqref{eq:B_compact} under the same interchange.

\section{Pairing terms}
\label{app:pairing_elements}

For a singly quantized vortex with on-site, sublattice-uniform pairing
$\Delta(\bm{r}) = \Delta(r)e^{-i\theta}$, the matrix elements between
Bessel basis states follow directly from the angular selection rule of
App.~\ref{app:bdg_matrix},
\begin{equation}
\langle jm|\Delta(\bm{r})|j',m'\rangle
= \delta_{m',m+1} C^{jj'}_{m,m+1} \tau_0,
\label{eq:pairing_matrix_element}
\end{equation}
where $\tau_0$ reflects the on-site, sublattice-diagonal form of
the pairing field, and the radial overlap is
\begin{equation}
C^{jj'}_{m,m+1}
\equiv \int_0^R  rdr
\Phi_{jm}(r)\Delta(r)\Phi_{j',m+1}(r).
\label{eq:C_pairing}
\end{equation}
Hence $\bm{\Delta}_{mm'} = \bm{0}$ unless $m'=m+1$, as already
anticipated in Eq.~\eqref{eq:delta_selection}. The conjugate sector
$\langle jm|\Delta^{*}(\bm{r})|j' m'\rangle$ similarly enforces
$m'=m-1$ and produces the overlap $(C^{j'j}_{m-1,m})^{*}$ entering
the hole equation. Together, the $m'=m+1$ block and its conjugate
$m'=m-1$ counterpart constitute the full pairing matrix $\bm{\Delta}$
and its Hermitian conjugate $\bm{\Delta}^\dagger$ appearing in
Eq.~\eqref{eq:bdg_full_compact}.

\section{Self-consistency relations}
\label{app:observables}

The self-consistency loop of Algorithm~1 is completed by expressing
the gap and density equations~\eqref{eq:gap}-\eqref{eq:density} in
terms of the Bessel basis coefficients $\{u^{S}_{njm}, v^{S}_{njm}\}$.
The key simplification is that every observable is a thermal average 
over the angular coordinate,
$
A(r) = \frac{1}{2\pi}\int_0^{2\pi}  d\theta A(\bm{r}),
$ 
so each angular integral reduces to a Kronecker delta via
Eq.~\eqref{eq:angular_orthogonality}, fixing the angular-momentum
bookkeeping in each case.

For the gap equation, substituting the
expansion~\eqref{eq:bessel_expansion_vector} into
Eq.~\eqref{eq:gap} and stripping the vortex phase
$\Delta(\bm{r})=e^{-i\theta}\Delta(r)$ leaves an angular factor
$e^{i(-m -1 +m')\theta}$ under the average, which
Eq.~\eqref{eq:angular_orthogonality} collapses to $m'=m+1$. The
radial gap profile then reads
\begin{align}
\Delta(r) = -\frac{U}{2\pi}\sum_{njj'm}
u^{S}_{njm} v^{S*}_{nj',m+1} 
\Phi_{jm}(r) \Phi_{j',m+1}(r) f(E_{n}).
\label{eq:gap_bessel}
\end{align}
The structure of Eq.~\eqref{eq:gap_bessel} is consistent with the
pairing block of the BdG matrix~\eqref{eq:delta_selection}: particle
components at angular momentum $m$ couple to hole components at $m+1$.
For the density equations, the spin-resolved densities are scalars,
so the angular average simply sets $m=m'$. Using
$f(-E_n) = 1-f(E_n)$ and inserting
Eq.~\eqref{eq:bessel_expansion_vector} into
Eq.~\eqref{eq:density} gives
\begin{align}
\rho_{\uparrow}(r)
&= \frac{1}{2\pi}\sum_{njj'mS}
u^{S*}_{nj'm} u^{S}_{njm}
\Phi_{jm}(r) \Phi_{j'm}(r) f(E_{n}),
\label{eq:rho_up_bessel}\\
\rho_{\downarrow}(r)
&= \frac{1}{2\pi}\sum_{njj'mS}
v^{S*}_{nj'm} v^{S}_{njm}
\Phi_{jm}(r) \Phi_{j'm}(r) f(-E_{n}),
\label{eq:rho_down_bessel}
\end{align}
where $f(-E_{n})=1-f(E_{n})$.
The total local density feeding the algorithm is
$\rho(r)=\rho_{\uparrow}(r)+\rho_{\downarrow}(r)$, summed over the
sublattice index. Equations~\eqref{eq:gap_bessel}-\eqref{eq:rho_down_bessel}
are the explicit basis representations evaluated at each iteration of
the self-consistent BdG solver, with the updated $\Delta(r)$ feeding
back into the radial overlaps $C^{jj'}_{m,m+1}$ of
Eq.~\eqref{eq:C_pairing}.

\bibliography{refs}

\end{document}